\documentstyle[preprint,aps,tighten]{revtex}
 \def\b{\bigskip}
 
 \def\n{\noindent}
 
 \def\be{\begin{equation}}
 \def\ee{\end{equation}}
 \def\bea{\begin{eqnarray}}
 \def\eea{\end{eqnarray}}
 \def\bec{\begin{center}}
 \def\enc{\end{center}}
 \def\nnu{\nonumber}
 \def\ba{\begin{array}}
 \def\ea{\end{array}}
 \def\AS#1{\renewcommand{\arraystretch}{#1}}

 \def\WW{Weisskopf--Wigner}
 \def\DE{Dyson equation}
 \def\BSE{Bethe--Salpeter equation}
 \def\RWA{Rotating wave approximation}
 \def\BA{Born approximation}
 \def\BZ{Brillouin zone}
 \def\SCBA{Self-consistent Born approximation}
 \def\SC{self-consistent}
 \def\GF{Green function}
 \def\NGF{non-equilibrium Green function}
 \def\VC{virtual crystal}
 \def\IC{initial condition}
 \def\se{self-energy}

 \def\Eq#1{(\ref{#1})}
 \def\EEq#1{Eq. (\ref{#1})}

 \def\VV{\rule{0em}{1.1em}}
 \def\DIA{$\diamond$}

 \def\ci{{\mbox i}}
 \def\ket#1{|#1\rangle}
 \def\bra#1{\langle#1|}

 \def\Alloy#1#2#3#4#5{{\rm #1}_{#2}{\rm #3}_{#4}{\rm #5}}

 \def\eps#1#2#3{{\epsilon}_{#1}^{#2}{#3}}
 \def\Om{\Omega}
 \def\Omi{{\mit \Omega}}
 
 \def\ep{\epsilon}

 \def\half{{\textstyle \frac{\scriptstyle 1}{\scriptstyle 2}}}
 \def\Tr{{\rm Tr}}
 \def\Re{{\rm Re}}
 \def\Im{{\rm Im}}
 \def\Abs#1{|#1|}
 \def\Ave#1{\langle#1\rangle}
 \def\ave#1{{\scriptstyle\langle}#1{\scriptstyle\rangle}}
 \def\Gm{{\bf G}}
 \def\Sim{{\bf\Sigma}}
 \def\Bfc#1{\mbox{\boldmath $\cal #1$}}
 \def\Bf#1{\mbox{\boldmath $#1$}}
 \def\Gr{G^R}
 \def\Ga{G^A}
 \def\Gl{G^<}
 \def\Gg{G^>}

 \def\Gcl{{\cal G}^<}

 \def\Si{{\mit \Sigma}}
 \def\Di{{\mit \Delta}}
 \def\Sir{{\mit \Sigma}^R}
 \def\Sia{{\mit \Sigma}^A}
 \def\Sil{{\mit \Sigma}^<}
 
 \def\ci{{\rm i}}
 \def\Dt{\ci\hbar\,\partial_t}
 
 \def\SSC#1{{\scriptstyle {\sc#1}}}
 \def\TRA#1{\stackrel{#1}{\longrightarrow}}
 
 \def\roo{\VV_{o\!}{\rm\rho}_{o}}
 \def\dlangle{\langle\!\langle}
 \def\drangle{\rangle\!\rangle}

\draft

\begin{document}

 \title{Photoexcited transients in disordered semiconductors:\\
 Quantum coherence at very short to intermediate times}

 \author{A. Kalvov\'{a}}
 \address{Institute of Physics, Academy of Sciences of the Czech
 Republic, Na Slovance 2, CZ-182 21 Praha 8, Czech Republic}
 \author{B. Velick\'{y}}
 \address{Institute of Physics, Academy of Sciences of the Czech
 Republic, Na Slovance 2, CZ-182 21 Praha 8, Czech Republic
 \\and \\Faculty of Mathematics and Physics, Charles University,
 Ke Karlovu 5, CZ-121 16 Praha 2, Czech Republic}


\maketitle
\nopagebreak
 \begin{abstract}

 We study theoretically electron transients in semiconductor alloys
 excited by  light pulses shorter than 100 femtoseconds
 and tuned above the absorption edge
 during and shortly after the pulse, when  disorder scattering
 is dominant.
 We use non-equilibrium Green functions employing the
 field-dependent self-consistent Born approximation. The
 propagators and the particle correlation function are obtained by
 a direct
 numerical solution of
  the Dyson equations in differential form.
  For the purely elastic scattering in our model system
  the solution procedures for the retarded propagator and for the
 correlation function can be decoupled.The propagator is used as
 an input in calculating the correlation function.

 Numerical results combined with a cumulant expansion
   permit
 to separate in a consistent fashion the dark and the induced parts of the
 self-energy. The dark behavior reduces to propagation of  strongly
 damped quasi-particles; the field induced self-energy leads to
 an additional time non-local coherence.

 The particle correlation function
 is formed by a coherent transient and an
 incoherent back-scattered component.
 The particle number is conserved only if
 the field induced  coherence
 is fully incorporated.
 The transient polarization and the energy balance
 are also obtained and interpreted.
\end{abstract}



 \newpage

 \section{Introduction}
 \label{introduction}

 The response of electrons in a semiconductor to a
 strong short optical pulse is an important case of transient non-equilibrium processes
 in many body systems. It is discussed in detail in monographs
 by Haug and Jauho\cite{book:Haug} and by Bonitz\cite{book:Bonitz}. Step by step,
 the technique of non-equilibrium Green's functions
 (NGF) has been recognized as the most reliable and physically most transparent
 method for handling strongly non-equilibrium processes. However, because of the
 prohibitive requirements on computational means, its practical use was until
 recently connected with various approximation schemes reducing the full
 Kadanoff-Baym equations (KBE)or their equivalents to simplified quantum
 transport equations by means of various decouplings ("ansatzes"). The generic
 ansatz appears already in the classical works by Kadanoff and
 Baym\cite{book:Kadanoff} and by Keldysh\cite{Keldysh}, and it was followed by a
 number of its refinements. This work was very successful in describing a wide
 variety of experimental data almost quantitatively. Still, uncertainties
 concerning the validity and more subtle properties of the additional
 approximations remain.

 Since early times, however, there existed an important line of research
 \cite{Danielewicz},\cite{Kohler} on the direct solution of the KBE for transients,
 both in electron systems and in nuclear matter, where it was mandatory
 to use the NGF and related quantities as two time functions.
 More recently,  efforts have concentrated precisely on  a direct
 solution of the full NGF incorporating both the electron-phonon and
 electron-electron interactions, as discussed already in Ref. \cite{book:Bonitz}
 and expounded in more detail in the proceedings\cite{book:Bonitz_ed}. The need
 to work with two time functions
 limits the physical approximations for the self-energies, as
 analyzed in the general references given above.

 We will consider the third principal mechanism of electron scattering, caused
 by a static random potential. In random semiconductor alloys the strongest
 scattering is often caused just by the disorder, which then dominates the early
 stages of the transient response to a pulsed disturbance. The case of disorder
 has some interesting special properties while providing an excellent model
 situation to study general aspects of non-linear transients
   \cite{we:zschrI}\cite{we:noeks4}\cite{we:zschrII}\cite{we:noeks5}.
   This is well
 illustrated on our two-band model semiconductor with disorder only acting in
 the conduction band\cite{we:pisa}. A dipole optical transition transfers a
 valence electron with a sharp wave vector value $k$ to the conduction band. For
 one alloy configuration, the conduction states are random and have no wave
 vector at all. In the one-electron picture, this is a fully unitary evolution
 in which the transition is made to many of the random states with a full
 preservation of the quantum coherence. Alternatively, a configuration average
 is performed\cite{book:Gonis} accounting for the fact that the macroscopic
 response of the electrons is self-averaging\cite{Kohn-Luttinger}. The static
 random fluctuations of the alloy potential play the role of the quantum field
 fluctuations of the many body theory ("frozen phonons"). Now, the transitions
 preserve the $k$-vector, but the final states are renormalized and
 decaying, independently of the excitation process or of the electron
 distribution. In analogy with the phonon case, this may be termed a dark
 polaron effect. The optical jump injects an electron into a bare Bloch
 conduction state, and its transition into the dressed state (formation process)
 plays an important role for the short time dynamics of the coherent excitation.
 At longer times, the incoherent back-scattering sets on to compensate for the
 decay of the coherent population. All these features coincide with those
 occurring in a genuine quantum field case, and it appears that for the disorder
 case the full NGF treatment is the only reliable approach. In paper
 \cite{we:zschrI}, we have developed the basic theory for NGF in the presence of
 external fields and alloy scattering and solved the equations of motion in
 the coherent potential approximation (CPA, for the equilibrium case and linear
 response see Ref. \cite{book:Gonis}) in a closed form. The solution was limited
 to an unrealistic case of a sharp onset of steady illumination, however. In
 \cite{we:noeks4}, this solution was used for a numerical test of one of the
 successful ansatzes, the so-called GKBA\cite{lipavsky}. In \cite{we:zschrII},
 an adiabatic approximation was developed for slowly varying pulses, employing
 the mixed time-energy representation, and in \cite{we:noeks5}, we were able to
 present early results for retarded propagators in the presence of pulses of a
  general shape.

 This paper extends our previous work
 to a full computation of the response to an arbitrary pulse,
 although only in the weak scattering limit of the CPA, the self-consistent Born
 approximation (SCBA).

 The alloy scattering on a static potential is specific in that it is strictly
 elastic and instantaneous, with no internal time structure. This has an
 important consequence. The electron propagators do not depend on the electron
 distribution, being basically a configuration averaged one-electron evolution
 operator. The propagators can thus be studied, analytically and numerically as
 well, in an independent and comparatively easy step.
  In the second step yielding the full NGF, the propagators serve as an input
  and it is sufficient to solve a single linear equation for, say, the particle
 correlation function, instead of the usual set of two
 coupled non-linear Kadanoff-Baym equations.

 On top of that, the use of this asymmetric set of Dyson
 equations \cite{LW} permits a direct interpretation of some features of the
 non-equilibrium process. By concentrating first on the propagators, we can
 study in detail the formation stage and the transition to the long time
 ("quasi-particle") regime. The self-energy can be separated into its dark part
 and the induced component, which reflects the coherent coupling between the
 excitation and the scattering. Proper inclusion of this coherence effect
 appears to be essential to keep the theory conserving and consistent.
 This becomes apparent in the behavior of the particle correlation
 function,  in the loss of coherence in the non-equilibrium electron distribution
 and in the validity  of conservation laws.

 We start, in Sec. \ref{sec:model}, by describing  chemical disorder in
 semiconductor alloys (\ref{subsec:alloys}) which serves to justify the simple
 two band model with a short-range disorder of
 atomic levels acting only in the conduction band, as introduced
 in \ref{subsec:two_band}.
 The effect of the alloy disorder in the dark is
 discussed in Sec. \ref{sec:dark_alloy}. First, the SCBA is introduced and the
 renormalized bands are obtained (\ref{subsec:dark_alloy_bands}). For an actual
 choice of the model parameters, the renormalized bands, their damping, and the
 weak field one-photon resonance conditions are presented
 (\ref{subsec:parametrization}). The next subsection \ref{subsec:gf_dark_time}
 deals with the dark propagators in the time domain. The formation process
 leading from a bare electron to a quasi-particle is analyzed. The SCBA is
 compared with the exact short time cumulant expansion of the GF and the
 formation time is estimated.

 In Sec. \ref{sec:resp_calc}, the action of the light pulse is included.
 Specific properties of NGF in
 disordered systems are introduced in \ref{sec:ngf}.
 General relations for the Langreth-Wilkins form of NGF are  in
 \ref{subsec:gen_rel}.
  Next two
 subsections \ref{subsec:single_conf}, \ref{subsec:conf_ave}  introduce
 the GF for a single alloy configuration
 and its configuration average. A closed form of the
 perturbation expansion is obtained.
 In \ref{subsec:work_dyson}, the explicit equations of motion for all components
 of the NGF are given (\ref{subsec:g_less_expl}).
 In  \ref{subsec:dark_sep} we separate the
 dark polaron effect from the  light induced self-energy. Its properties are analyzed
 in  \ref{subsec:ind_se},
 where the short time expansion is extended also to the excited system.
 The meaning of the dark quasi-particle
 approximation is clarified in  \ref{subsec:dark_qp}.
 The actual computational schemes for both
 Dyson equations are presented in Subsec. \ref{subsec:comp_scheme}.

 Sec. \ref{sec:numerical} is devoted to numerical results.
 The actual choice
 of the pulse is specified (Subsec. \ref{subsec:pulse_par}).
 The retarded propagator is presented in \ref{subsubsec:g_reta}.
 An example of the induced self-energy is shown in Subsec.
 \ref{subsec:calc_sir} and its role   for the propagator is demonstrated
 in \ref{subsec:calc_gr}. The related  coherence and time non-locality leads to
 a breakdown of the semi-group property of the propagator, Subsec.
 \ref{subsec:test_semi}.
 The computed behavior of the particle
 correlation function is described in \ref{subsec:g_less}, presenting
 $\Sigma^<(t,t')$ in \ref{subsubsec:part_se} and an example of
 $G^<({\rm k};t,t')$ itself in \ref{subsubsec:part_co}.

  Physically, we are particularly 
 interested in the averaged one-particle density matrix, and in
 the observable averages computed with its aid.
  Sec. \ref{sec:physical_prop}  is divided into
 two subsections.
 In the first one, \ref{subsec:rho_and_obs},
 we consider the
 connection between the one electron density matrix and the particle
 correlation function, construction of the precursor kinetic equation
 and a proof of the particle number conservation in SCBA
 (\ref{subsubsec:density_matrix}). Several observables and their average
 values   are treated in
 \ref{subsubsec:obs_and_ave}.
 In the last subsection
 \ref{subsec:computed_rho} we show the results of the corresponding
 computations. First, individual matrix elements of $\varrho(t)$ are shown and
 discussed in \ref{subsubsec:distr_f}.
 The last part \ref{subsubsec:total} concerns observables. The light-disorder coherent
 coupling conditions the particle number conservation which hinges
 upon a full inclusion of the induced \se\ in the computation. The
 energy is not dissipated in the model, only transferred between the
 pulse and the electrons, while the dephasing of the interband
 offdiagonal components of the polarization need not be introduced
 by hand, as it appears to be linked with pronounced cancellations of
 the oscillatory integrand in the trace yielding the total
 polarization.

 Two appendices outline the steps necessary to incorporate into the
 theory random initial conditions (\ref{app:corr_ic}) and observables represented by random
 operators (\ref{app:rand_obs}).

 \section{Model description of the system}\label{sec:model}

 In this paper, we use the model semiconductor alloy introduced previously \cite{we:zschrI},
 \cite{we:zschrII}. The simplifications inherent to the model bring it to a
 close parallel to two-band models with electron--phonon interaction
 \cite{book:Haug}. Static alloy disorder in this picture may be understood in
 terms of "frozen phonons". The model belongs to the family of "independent band
 models" used to study optical transport in disordered systems \cite{we:pisa}.

 \subsection{Electrons in semiconductor alloys}\label{subsec:alloys}
 One aim of this paper is to turn attention
 to the importance of disorder effects in the optical response to short pulses;
 once the excitation will take place at regions of the band structure even moderately
 beyond the absorption edge, the alloy scattering in a mixed crystal
 will frequently become
 stronger than either the $e-e$ or the $e-p$ collisions
 and it will represent
  the principal mechanism of relaxation of the photoexcited population during
 the initial period. As an example, consider the zinc blende semiconductors.
 Even for a conservative choice,
  the $\Alloy{Ga}{1-c}{Al}{c}{As}$ system on
 the GaAs side,  $c\leq$ 0.1,
 pronounced alloy scattering effects are predicted\cite{Chen1981}. Stronger
 effects still can be expected for other systems, in particular some of the
 II-- VI alloys\cite{hass1984}. In all such materials, the
 basic alloying mechanism is an  {\em isoelectronic substitution}, by atoms from
 the same group of the periodic system. As a consequence, the random  one-electron
 potential in the alloy is dominated by short range, basically site--localized
 fluctuations. In an orbital picture, this is best captured by a one-electron
 Hamiltonian in a localized orbital ("tight binding") representation \cite{book:Gonis},
 which is
 by now standard, and considered quantitative \cite{book:Harrison}, \cite{book:Cardona}.
 The tight binding Hamiltonian in the minimum basis of quasiatomic orbitals will
 typically involve (i)~non-random off-diagonal hopping matrix elements responsible
 for bonding between the $sp^3$ hybrids, and for creation of the corresponding
 hybridization gap between the valence and the conduction bands, and (ii)~random
 diagonal elements, quasiatomic levels $\ep_{\ell}$, with as many admissible values, as
 are the alloy components (see, for example,
 \cite{hass1984}).In the case of $\Alloy{Ga}{1-c}{Al}{c}{As}$, the doping proceeds only on
 the cationic
 sublattice. In the Harrison parametrization\cite{book:Harrison}:
 \bec
 $\eps{4s}{\rm Ga}=-11.37\,{\rm eV},\ \eps{3s}{\rm Al}=-10.11\,{\rm eV}$\\
 $\eps{4p}{\rm Ga}=-\,4.90\,{\rm eV},\ \eps{3p}{\rm Al}=-\,4.86\,{\rm eV}$. \enc
 If the alloy is {\em random}, that is, without any short or long range order,
 the random potential is fully characterized by the $\ep$ values and the
 concentration $c$.
 In particular, the
 ratios {\sc (level difference)/bandwidth} determine the alloy regime.
 If they are much less than one (as is the case for $\Alloy{Ga}{1-c}{Al}{c}{As}$),
 a perturbation approach is justified. The $\ell$-contribution to the level
 broadening  is then proportional to the projected density of states of a
 given symmetry.
 In general, spectral weight
 of each orbital is divided between the conduction bands and the
 valence bands, so that, say, the disorder of the cation $s$ levels affects both.
 The projected spectral weights are distributed unevenly, however. Around the
 gap, where the optical excitation is assumed  to act, the conduction band is
 predominantly composed of cation states, and these mostly with the
 $s$-symmetry. The top of the valence band, on the contrary, is nearly  anion $p$
 by nature. The cationic disorder, as a result, acts much more strongly at the
 bottom of the conduction band than at the top of the valence band.

  \subsection{Two band model}\label{subsec:two_band}
  On the basis of
 these remarks, we will now develop a simplified model of the  electronic
 structure of this system as used  in Refs.
 \cite{we:zschrI}, \cite{we:zschrII}.
 \begin{enumerate}
 \item We consider a two-band semiconductor with the gap between
 two isotropic parabolic band edges at the center of the Brillouin zone
 ({\em "standard band structure"})
 \item All many-body interactions are ignored
 \end{enumerate}
  In general, there are three mechanisms by which the disorder couples both
 bands:
  the $c-v$ chemical
 hybridization mixing the conduction and the valence states,  a statistical
 correlation between the electrons moving in both bands caused by the  atomic
 disorder configuration they have in common, and a dynamical transfer of disorder
 between the bands
 due to the mixing of optically coupled valence and conduction states. We will
 eliminate both static coupling mechanisms by hypothesis.
 \begin{enumerate}
 \setcounter{enumi}{2}
 \item The disorder acts in each band separately, not mixing states of both
 bands
 ({\em "independent bands"})
 \item The disorder in the valence band is neglected ({\em "cationic doping"}).
 \item The effect of the optical pulse is restricted to a non-random interband
 dipole coupling treated in the rotating wave approximation (RWA).
 \end{enumerate}
 This model appears as reasonable for an optical excitation slightly beyond the
 absorption edge,  but the band structure  has to be extended
 to the whole \BZ\ for two reasons:
  \  \DIA\ the single site
 nature of the disorder spreads its effect homogeneously over the whole \BZ,
  \DIA\  a short strong optical pulse acts
 in an extended part of the \BZ, ($E - t$ uncertainty).
 This becomes apparent, for example, in the band renormalization
 caused by the disorder, and, in particular, in the short time quasi-particle
 formation process.
 The band parts remote from the \BZ\ center can be
 described rather schematically, however, as only their gross properties enter the
 problem.

 Disorder does not mix the two bands.
 The band projectors of these "independent bands"
 are non-random and diagonal in both the Bloch and the Wannier
 (site)
 basis:
 \be
    P_b =\sum\limits_{\sf BZ}\ket{b{\rm k}} \langle b{\rm k}| =
    \sum\limits_{\rm lattice}|bi\rangle \langle bi|
  \ , \ \ b=c,v . \label{eq:projectors}
 \ee
 The full one-electron Hamiltonian for one  configuration of an
 $\Alloy{A}{1-c\,}{B}{c\,}{}$ alloy has the
 following structure:
 \be
   {\cal H}=W_v +W_c +{\cal V}_c + U(t) .
 \label{eq:hamiltonian}
 \ee
 The  non-random configuration
 independent or averaged quantities are denoted by italics, while the configuration
 dependent ones by rounded ("\verb= \cal =") characters.

 Here, $W_v +W_c$ is the Bloch Hamiltonian of the A crystal. Both band Hamiltonians
 are diagonal in the Bloch basis:
 \be W_b= P_b W_b  P_b = \sum_k|bk\rangle \epsilon_b (k)
 \langle bk|, \qquad b=c,v
 \label{eq:bands}
 \ee
 The random potential acts only in the conduction band and it is site diagonal:
 \AS{1.7}
 \be
 \ba{ccrcccc}
  {\cal V}_c&=&  P_c {\cal V}_c  P_c&=&\sum\limits_i |ci\rangle \epsilon
  _i\langle ci|
   \label{eq:rand_pot}
  \ea\ee
  Here, $\ep_i$ are quasiatomic level shifts $\ep^A\ {\rm or}\ \ep^B$.

 Finally, $U(t)$ is a non-random inter-band dipole coupling of the
 pulse to the electrons, whose electric field
 is a linearly polarized harmonic wave with a basic frequency
 $\Omega$ and an envelope ${\bf E}_\SSC{m}\Phi(t)$ with $\Phi\geq 0,\ \max\Phi=1$
 ,${\bf E}_\SSC{m}$ is the peak value of the field:
 \be
 {\bf E}(t)={\bf E}_\SSC{m}\cos(\Om t)\Phi(t){\bf e}
 \label{eq:ext_field}
 \ee
 Assuming validity of the \RWA\ (RWA), and making at the same time the approximation
 of a $k$-independent transition matrix element, we get for $U(t)$:
 \be\ba{rcl}
  U(t)&=&U_{cv}(t)+U_{vc}(t)=-e({\bf r}_{cv}+{\bf r}_{vc}){\bf
  E}(t) \\
  &\TRA{\sc rwa}&-\Phi(t)\left\{
  \sum\limits_k\ket{c{\rm k}}Q{\rm e}^{-\ci\Om t}\bra{v{\rm k}}
  +\sum\limits_k\ket{v{\rm k}}Q^\ast{\rm e}^{\ci\Om t}\bra{c{\rm k}}\right\}\\
  &&{\rm r}_{cv} = P_c{\rm r}P_v,\ {\rm etc.},
  \quad Q = \frac{1}{2}e{\rm E}_\SSC{m}\bra{c{\rm k}\!\simeq\! 0}
  ({\bf e},{\bf r})\ket{v{\rm k}\!\simeq\! 0}
  \label{eq:interband_term}
 \ea\ee
 The coupling strength parameter $Q$ can  be made real positive, $Q=Q^\ast>0$
 by rephasing. RWA allows to make a
 time-dependent unitary transformation
 \be
  O =P_c+P_v{\rm e}^{-\ci\Om t}  \label{eq:gali}
  \ee of the
 Hamiltonian (the Galitskii transformation \cite{galitskii})
 to eliminate the rapid $\Om$ oscillations of $U(t)$.
 Thus, ${\cal H}(t)\TRA{O}\tilde{\cal H}=O{\cal H}O^\dag-O\Dt O^\dag$:
 \be\ba{rcl}
 \tilde{\cal H}(t)&=&\tilde{W}_v +{W}_c +{\cal V}_c  + \tilde{U}(t)\\
 &&\tilde{\ep}_v({\rm k})={\ep}_v({\rm k})+\hbar\Om
 \\
 \tilde{U}(t)&=&-Q\Phi(t)\left\{
  \sum\limits_k\ket{c{\rm k}}\bra{v{\rm k}}
  +\sum\limits_k\ket{v{\rm k}}\bra{c{\rm k}}\right\}
 \label{eq:transf_ham}
 \ea\ee
 The valence band is displaced  by $\hbar\Om$ upwards on the transformation.
 The interaction with light appears now as a transient hybridization
 of the transformed bands giving rise to the band splitting sometimes called
 the "Galitskii gap".

 {\em From now on, we will work in the Galitskii picture dropping
 the tilde accents over the operators}, but we will keep the tildes to indicate
 the shifted valence band energies.

 Denoting the usual configuration average by $\Ave{\,\dots\,}$,
 we may introduce the {\em mean-field Hamiltonian} $H_{\SSC{mf}}$
 and the configuration dependent {\em random field} ${\cal D}$:
 \bea
 {\cal H}(t)&=&H_{\SSC{mf}}(t)+{\cal D}(t)
 \label{eq:split_mf}\\
 H_{\SSC{mf}}(t)&=&\Ave{{\cal H}(t)} \label{eq:def_mf}\\
 \Ave{{\cal D}(t)}&=&0 \label{eq:ave_d}
 \eea
 In our model, ${U}(t)$ is non-random, while ${\cal V}$
 is time independent. Forming $H_\SSC{mf}$ then means that
 the {\em \VC\ }part of ${\cal V}_c$ responsible for the rigid band shift
  is transferred to the renormalized conduction band  $\overline{W}_c$:
 \be
   H_\SSC{mf}={W}_v+\overline{W}_c +{U}(t)
  \label{eq:hmf_expl}
 \ee
 with
 \be
 \ba{rcl}
  \overline{W}_c&=&\sum\limits_k|ck\rangle \overbrace{
 {\ep}_c({\rm k})+\ave{\ep}}^{\displaystyle {\ep}_\SSC{mf}({\rm k})}
 \langle {\it ck}|\\
 {\cal D}&=&  \sum\limits_i|ci\rangle \underbrace{
 (\epsilon _i-\ave{\epsilon})}_{\displaystyle d_i}
 \langle {\it ci}|
 \ea \label{eq:rand_fluct}
 \ee
  where $\ave{\ep}$  is the average atomic energy
 $(1-c)\ep^A+c\ep^B$. The A crystal will be doped with B atoms,
 thus we choose
 \be
  \ep^A=0 ,\ \ \ep^B=\delta,\ \ \ave{\ep}=c\cdot\delta
 \label{eq:def_delta}
 \ee
 There are thus two alloy disorder parameters, concentration $c$
 and the level fluctuation $\delta$.

 \section{Dark alloy}\label{sec:dark_alloy}
 The disorder scattering is present even without  the optical excitation, and
 the $c$-band states are correspondingly distorted. The excitation transfers
 electrons into these {\em disorder dressed} states, and the behavior of the
 light induced transition states can best be described differentially with
 respect to the dark dressed states. This situation is similar to the dark
 polaron effect in the initial state \cite{haugbutwhich}.
 The poles of the renormalized dark alloy bands define the weak field resonance
 transitions, but for strong pulses, the short time deviations from the pole
 (quasi-particle) behavior are essential.

In this section, we will consider  this dark alloy  case. This is an equilibrium
situation 
 characterized by a time independent ${\cal H}$.
 While in general the electron response is conveniently expressed in terms of the
 \NGF, in equilibrium
 it is sufficient to consider the retarded propagator $\Gr$ of one electron moving
 in the dark conduction band.

 The propagator then
  depends just on the time difference $t-t'$, and may be analyzed without invoking
  the NGF formalism in the
 spectral representation introduced by Fourier transformation:
 \AS{1.3}\be
 \Gr({\rm k},t)=
 \displaystyle\int{{{\rm d}E}\over{2\pi\hbar}}\,
 G({\rm k},E+\ci 0)\cdot{\rm e}^{-\ci\,E\,t/\hbar}
 \label{eq:gr_fou}
 \ee

 \subsection{Dark alloy bands in SCBA}
 \label{subsec:dark_alloy_bands}
  A single conduction band with diagonal disorder is easy for  treatment
 in any single site approximation, as described in detail in \cite{book:Gonis}.
 Here, we only sketch the \SCBA\ results. The \GF\ in the energy representation
 coincides with the configuration averaged resolvent, and is given by
 \be
 G({\rm k},z)={1 \over {z-\ep_k-\ave{\ep}-\Si(z)}}
 \label{eq:gf_dark}
 \ee
 Here, $z$ denotes the complex energy, $\ep_k\equiv{\ep}_c({\rm k})$ and
 the SCBA self-energy does not depend on ${\rm k}$. It is determined from
 \be \ba{rcl}
 \Si(z)&=&\gamma\cdot F(z)\\
 F(z)&=&N^{-1}\sum\limits_kG(k,z)\\
 \gamma&=&c(1-c)\delta^2
 \label{eq:scba_dark}
 \ea\ee
 The Nordheim parameter $\gamma$
 measures the scattering strength of the random potential, $F$
 is the local \GF (LGF); its imaginary part determines the conduction band DOS per cell
 by $g_c(E)=\mp\pi^{-1}\Im F(E\pm\ci 0)$. Combining Eqs. \Eq{eq:gf_dark} and
 \Eq{eq:scba_dark}, we express this local GF in the alloy in terms of $F_o$,
 the pure A crystal LGF, with a shifted argument:
 \bea
 F(z)&=&F_o(z-\ave{\ep}-\Si(z))
 \label{eq:reduction}
 \eea
 The SCBA equation \Eq{eq:scba_dark} reads explictly:
 \bea
 \Sigma(z)&=&\gamma\cdot F_o(z-\ave{\ep}-\Si(z))
 \label{eq:scba_expl}
 \eea

 This can be solved by iteration.
 The dispersion law and the pole approximation, however, may be obtained explicitly.
 Let $\delta$ grow from zero to its actual value. The band
 energies turn adiabatically into the complex resonance energies,
 $\ep_k\equiv{\ep}_c({\rm k})\mapsto z_k\equiv{z}_c({\rm k})$,
 given by the poles of the \GF\ \Eq{eq:gf_dark} (on the analytic continuation
 to the non-physical sheet of the $z$-Riemann surface]:
 \be \ba{rcl}
 G^{-1}({\rm k},z_k)&=&0\\
 z_k&=&\ep_k+\ave{\ep}+\Si(z_k)
 \label{eq:pole_condition}
 \ea\ee
 This  holds for any $k$-independent \se. In the SCBA,
 \Eq{eq:pole_condition} becomes an explicit expression for the
 SCBA dispersion law,
 \be \ba{rcl}
  z_k&=&\ep_k+\ave{\ep}+\gamma\cdot F_o(\ep_k+\ci 0)
 \label{eq:disp_law}
 \ea \ee
 The result looks like a most naive use of the non-selfconsistent \BA.

 Next, we rewrite
 \EEq{eq:gf_dark} into a residuum form,
 \be
 G(k,z)={Z_k \over {\,(z-z_k)\,}}+R(k,z)
 \label{eq:residual_form}
 \ee
 where $Z_k$ is the renormalization constant and $R(k,z)$ is regular
 in the neighborhood of the pole. From
 \Eq{eq:scba_expl} and \Eq{eq:pole_condition},
 we get also the renormalization constant explicitly:
 \be
 Z_k\equiv{1 \over {1-  \left.{{\rm d} \over {{\rm d}z}}
 \Sigma(z)\right|_{z_k}}}=1+\gamma\cdot\left. {{\rm d} \over {{\rm d}z}}
 F_o(z)\right|_{\ep_k }
 \label{eq:ren_const}
 \ee

 \subsection{Parametrization of the dark alloy model}
 \label{subsec:parametrization}

 A sufficient input for the dark
 alloy includes\ \ \DIA\ alloy parameters,
 $\delta$ and $c$,\ \ \DIA\ the $c$-band density of states per cell, $g_o$,
 from which the
 local GF $F_o$ can be obtained by the usual spectral representation
 \be
 F_o(z)=\int\,{\rm d} {\eta}\,{1 \over {\,z-\eta\,}}g_o(\eta)
 \label{eq:sp_repres} \ee
 Because we will be interested mostly in the spectral region of
 the lower band edge, a model $g_o$ may be considered satisfactory, if it has
 the correct energy position of this edge and the curvature yielding the
 required (density) effective mass. Otherwise, it should have  a proper
 bandwidth and it must satisfy the sum rule $\int\,{\rm d} {\eta}\,g_o(\eta)=1$.
 One easy way, adopted presently, to adjust the DOS to these requirements is to
 write it as an expansion in terms of the Tschebyshev polynomials of the second
 kind. An example of $F_o$ resulting from such procedure is shown in
 Fig.\ref{fig:f_o}. Details are given in an appendix to \cite{we:zschrII}.
 The parameters used were the band edge $\ep_c({\rm k}=\Gamma)\equiv
 E_G^A=1.5$ eV, the effective mass $m^\times_c=0.4\,m_e$,  the half-bandwidth
 $w_c=6\,$eV.

  The renormalized dispersion law $z_k=z_c({\rm k})$
  as a function of $\ep_k$, is given by \Eq{eq:disp_law}.
   The renormalization has two parts: a rigid shift $\ave{\ep}=c\delta$ of the whole
 band proportional to the alloy concentration, in agreement with the known
 notion of the {\em rigid band model} \cite{book:Gonis}, and a complex
 component $\gamma\cdot F_o(\ep_k+\ci 0)$, whose real part  gives the
 negative "polaron shift", and whose
 imaginary part is related with the electron life time by
 $\tau\approx\hbar/2\Abs{\Im z_k}$. The polaron shift varies with concentration
 in the "Nordheim" fashion, proportionally to $c(1-c)$. It is not entirely
 rigid, which leads to an energy and concentration dependent renormalization of
 the effective mass. This effect is minor in our case, however.

 In Fig. \ref{fig:z_k}, we display  by thick lines  the renormalized quasi-particle
 energies $z_k$  as a function of the bare band energy $\ep_k$, taking
 $\delta=0.84\,$eV, and varying $c$ in the range $0\div 0.5$. The energy dependent
 band broadening is depicted in Fig. \ref{fig:z_k} by lining each of the
 renormalized dispersion laws $\Re z_k$ by thin lines at a distance $\pm \Im
 z_k$. The rigid band energies are shown by dashed lines, and the polaron shift
 is given by the vertical distance between the full and the dashed lines.

  The corresponding renormalization constant is obtained from
 \Eq{eq:ren_const} and is shown in Fig.\ref{fig:Z_k}. This quantity is complex,
 and its module appears to be greater than 1, while the phase shift is negative.
 Thus, the long time behavior of the quasi-particle state will have the
 appearance of the simple \WW\ decaying state lagging somewhat behind in time.

  Returning to Fig. \ref{fig:z_k}, we discuss the plots (-.-.-) of the shifted
 valence band. With the basic frequency of the pulse specified,the valence band
 is shifted to $\tilde{\ep}_v({\rm k})={\ep}_v({\rm k})+\hbar\Om$. Its
 crossing with the renormalized conduction band is the point of the weak field
 one-photon resonance (vertical transition):
 \be
 z_k^\times\equiv {z}_c({\rm k}^\times)={\ep}_v({\rm k}^\times)
  +\hbar\Om\
  \label{eq:sc_crossing}
  \ee
 In the parabolic region, the two  bare bands are related by
  \be
  \ep_v({\rm k})={m_c \over m_v}(E_G^A-\ep_c({\rm k}))
  \label{eq:prop_bands} \ee
  (For simplicity, we will
 extend this as a definition for the entire model valence band.) Choosing
 $m_v=0.6m_e$ and taking first $\hbar\Om=2\,$eV, we plot in  Fig. \ref{fig:z_k}
 the corresponding line. The crossing points with the renormalized $c$-bands
 define the electron and the hole energy at resonance, and also the respective
 excess energies easy to read as the separations between the crossing point and
 the band edges at the vertical axis. The bare  energy of the crossing point
 serves to specify the position ${\rm k}^\times$ of the resonance in the
 Brillouin zone.

   The crossing points and the excess energies
 vary with the alloy concentration for a fixed $\Om$, and it would be preferable
 to adjust the basic frequency used, so that either the crossing point
 ${\rm k}^\times$ or the conduction band excess energy  be kept fixed. The
 vertical line laid through the $c=0$ crossing point defines, how the frequency
 should be varied with concentration in order to excite resonantly always the
 same part of the \BZ. In practice, a nearly equivalent way of tuning the
 frequency $\Om\rightarrow\Om^\times(c)$ may be to keep the excess energy equal
 to the difference $\hbar\Om^\times(c) -E_G(c)$ constant, at least in the \VC\
 sense.
 The numbers corresponding to Fig. \ref{fig:z_k} are listed in
 Tab. \ref{tab:1} and give an estimate of the importance of the
 alloy effects:
 \b\n It is seen that even the rather moderate alloy parameters produce a strong
 effect,
 in particular for the electron lifetimes. For stronger alloy scattering, the
 quasi-particle concept becomes useless, because it would be damped before
 completion of its formation.

 \subsection{Time dependence of the dark \GF}
 \label{subsec:gf_dark_time}

 Inserting the pole representation \Eq{eq:residual_form}into the \EEq{eq:gr_fou},
 we have
 \AS{1.3}\be
 \Gr({\rm k},t)={{\textstyle 1} \over {\textstyle \ci\hbar}}Z_k\,{\rm e}^
 {-{\scriptstyle \ci \over \scriptstyle \hbar}z_kt}+
 \displaystyle\int{{{\rm d}E}\over{2\pi\hbar}}\,
 R({\rm k},E+\ci 0)\cdot{\rm e}^{-\ci\,E\,t/\hbar}
 \label{eq:gr_pole}
 \ee
 The first term  prevails under typical conditions at sufficiently long times,
 and this corresponds to the  quasi-particle behavior.  

 To study arbitrary times, it is convenient to use an 'interaction picture' for
 $\Gr$ by factorizing off the mean field (\VC) propagator.
  For every ${\rm k}$,
 we define ({\em cf.} \Eq{eq:rand_fluct})
 \AS{2.5}\bea
 \ep_{\sc mf}&=&\ep_k+\ave{\ep}\\
 \Gr({\rm k},t)
 &\equiv&{{\textstyle 1} \over {\textstyle \ci\hbar}}{\rm e}^
 {-{\scriptstyle \ci \over \scriptstyle \hbar}{\scriptstyle \ep_{\sc mf}}\,t}\times
 {\rm e}^
 {-{\scriptstyle \ci \over \scriptstyle \hbar}\varsigma(t)}
 \label{eq:gf_dark_def}
 \eea
 The additional factor ${\rm e}^
 {-{\scriptstyle \ci \over \scriptstyle \hbar}\varsigma(t)}$
 describes the transformation to a quasi-particle state and its further
 evolution. We express it as a {\em complex phase factor} by introducing an "action"
 $\varsigma=\Re\varsigma+\ci\,\Im\varsigma$. Its real part describes the
 additional phase, the imaginary part measures the decay  of the amplitude
 $\Abs{\Gr}$.
 This identification is verified in the long time limit. By
 \Eq{eq:gr_pole}, the leading terms in $\varsigma$ represent the  "polaron shift"
 $z'_k-\ep_{\sc mf}$ and the damping $\Abs{z''}/\hbar$,
 \AS{1.3}\be
 \varsigma(t)\approx (z'_k-\ep_{\sc mf})t+\ci z''_k\,t+{\scriptstyle
 \cal O}(t)
  \label{eq:phase_pole}
 \ee
 For short times,  we may   employ the cumulant expansion of $G$, described in
 \cite{book:Brout}, p.3,\cite{book:Mahan}, Ch.2.  that is to expand in powers of time:
 \AS{2.5}\bea
 {\rm e}^
 {-{\scriptstyle \ci \over \scriptstyle \hbar}\varsigma(t)}&=&
 \sum\limits_{p=0}^{\infty}{{(-\ci)^p\,t^p} \over {\,p!\,\hbar^{p}\,}}\,M_p\\
 \varsigma(t)&=&\ci\hbar\sum\limits_{p=1}^{\infty}{{(-\ci)^p\,t^p} \over
 {\,p!\,\hbar^{p}\,}}\,C_p
 \label{eq:expansion_cumulant} \eea
  Here,
  $M_p$ are {\em  moments of the spectral density} of the state $\ket{{\bf
 k}}$ calculated with respect to $\ep_{\sc mf}$, while $C_p$ are the
 corresponding {\em cumulants}. The moments, at least for small $p$, can be
 calculated directly without difficulty, proceeding, for example,  similarly as
 in \cite{vke1968}. There are two options. Either the moments are calculated
 exactly, employing the fact that $\Gr$ coincides with the configuration
 averaged evolution operator up to a $1/\ci\hbar$ factor:
 \AS{2.5}\bea
 \Gr({\rm k},t) &=&{{\textstyle 1} \over {\textstyle \ci\hbar}}\bra{{\rm
 k}}\Ave{ {\rm e}^{-{\scriptstyle \ci \over \scriptstyle \hbar}{\cal
 H}\,t}}\ket{{\rm k}}\\
 &=&{{\textstyle 1} \over {\textstyle \ci\hbar}}{\rm
 e}^ {-{\scriptstyle \ci \over \scriptstyle \hbar}{\scriptstyle \ep_{\sc
 mf}}\,t}\times \bra{{\rm k}}\Ave{ {\rm e}^{-{\scriptstyle \ci \over
 \scriptstyle \hbar}({\cal H}-\ep_{\sc mf}\,1_{\rm o\!p})\,t}}\ket{{\rm
 k}}\\
 &=&{{\textstyle 1} \over {\textstyle \ci\hbar}}{\rm e}^ {-{\scriptstyle
 \ci \over \scriptstyle \hbar}{\scriptstyle \ep_{\sc mf}}\,t}\times
 \sum\limits_{p=0}^{\infty}{{(-\ci)^p\,t^p} \over {\,p!\,\hbar^{p}\,}}
 \underbrace{ \bra{{\rm k}}\Ave{({\cal H}-\ep_{\sc mf}\,1_{\rm
 op})^p}\ket{{\rm k}}}_ {\textstyle M_p}
 \label{eq:gf_dark_ex} \eea
 Or  the moments are
 determined within the SCBA. In that case, they should be identified with the
 coefficients of the Laurent expansion  of $G(z)$ at infinity,
 \be
 G({\rm k},z)=\sum\limits_{p=0}^{\infty}{{M_p} \over {\,(z-\ep_{\sc mf})^p}}
 \label{eq:gf_dark_Laur}
 \ee
 Now, we have to combine  Eqs. \Eq{eq:gf_dark}, \Eq{eq:scba_expl} and
 \Eq{eq:sp_repres}, using Laurent expansions of all quantities involved and
 determining the coefficients successively.

 Once the moments are determined in either way, the cumulants are obtained by
 comparing  both series in \Eq{eq:expansion_cumulant}. The results for $p=0\div4$
 are in the  table \ref{tab:2}; everything is expressed in terms of the moments
 of the VC density of states defined by
 $\mu_p^\SSC{vc}(E)=\int{\rm d}\bar{E}\,g_o(\bar{E}-\ave{\ep})(\bar{E}-E)^p$
 and of the alloy
 characteristics $c_a-c_B,\ \delta,\ \gamma$. It should be noted that because of
 the choice  $E=\ep_{\sc mf}$ as the reference energy, the first moment
 and cumulant both vanish, $M_1=C_1=0$.  The vanishing first moment simplifies
 the cumulant expressions in all orders:
  Up to $p=2$, the SCBA results and the exact ones are the same. Differences
 occur for higher $p$. The SCBA expressions use $\gamma$ as the only
 characteristic of disorder beyond the mean field, as implied by
 \Eq{eq:scba_dark}. In both approaches, the moments coincide with the cumulants
 up to the third order, and the first deviations occur in the fourth order as
 marked in boldface.

 We plot an example of the time evolution of the "phase" in
 Fig.\ref{fig:varsigma}. The gradual  onset of $\varsigma$ for the  shortest
 times justifies the conjecture that the one electron excitation starts as a
 bare (mean field) particle and develops the cloud of "frozen phonons" only in
 the course of time:
  \AS{2.5}\bea
  \Gr({\rm k},t)&=&{{\textstyle 1} \over
 {\textstyle \ci\hbar}}{\rm e}^ {-{\scriptstyle \ci \over \scriptstyle
 \hbar}\,t\,({\scriptstyle \ep_{\sc mf}} -{\scriptstyle 1 \over {\scriptstyle
 \;6\hbar^2}}C_3\cdot t^2+\ldots)} \times {\rm e}^ {-{\scriptstyle 1 \over
 {\scriptstyle \;2\hbar^2}}\gamma\cdot t^2+\ldots}
 \label{eq:short_time}
 \eea
 This behavior prevails
 only for very short times, and it is succeeded by a gradual transformation to
 the quasi-particle mode, as seen from the  figure. The characteristic time for
 this quasi-particle formation has the order $\hbar/(\ep_k-E_G)$. The energy
 entering this estimate is simply the energy distance from the nearest critical
 point of the density of states. In the present case, it yields a few, like 3,
 femtoseconds. Over this basic formative process, there is superimposed a much
 faster and rapidly damped evolution, with a characteristic time on the order of
 $10^{-1}$ fs. We enclose these details into small windows which are then blown
 up as inserts. The exact curves in the inserts are fitted by the lowest order
 cumulant approximations rather well. This testifies explicitly that the gross
 features of the whole band, as reflected in the lowest moments of the DOS are
 decisive here. In particular, it is interesting to inspect Fig.
 \ref{fig:varsigma}b depicting $\Im\varsigma$ related to $\Abs{G^R}$. Because
 $\Abs{Z}>1$, the quasi-particle GF starts {\em above} the true one.
 Nevertheless, the latter quantity starts by a rapid {\em decrease} due to a
 spreading of the probability amplitude into the whole band. Only in the second
 stage of the formation, a recuperation takes place, and the exact
 $\Im\varsigma$ starts to oscillate around the quasi-particle straight line. It
 is thus not easy in the present case to construct an interpolation form for the
 GF similar to that introduced in \cite{bhg1998}. Instead, we proceed as follows.

 The averaged dark GF obeys the \DE\ in the differential form,
 \be
 \Dt\Gr(t)=\ep_{\sc mf}\,\Gr(t)+\int{\rm d}\bar{t}\Sir(t-\bar{t})\Gr(\bar t)
 \label{eq:DE_dark_diff} \ee
   Instead of this
  integral term involving the \se, we may define, in analogy to the \WW\
  approximation, a multiplicative, single-time but $k-$dependent, quantity
  $\sigma_k(t)$ by
  \be
  \Dt\Gr(t)=\ep_{\sc mf}\,\Gr(t)+\sigma_k(t)\Gr(t)
  \label{eq:def_sigma} \ee
  An explicit "phase" integral equation relating the phase $\varsigma$ with
  $\Sir$ may be developed, but we will not pursue this task.

  Comparing with
  \Eq{eq:gf_dark_def}, we get the relation
  \be
  \sigma_k(t)={{\rm d} \over {{\rm d}t}}\,\varsigma(t)
  \label{eq:sigma_varsigma}
  \ee
  The last three equations combined will be used in Sec.\ref{subsec:dark_qp}
  to separate the dark and the induced parts of the \se\ in an efficient manner.
  This will be based on the short time behavior of $\sigma_k(t)$. As plotted
  in the other panels inserted in Figs. \ref{fig:varsigma}a, b, this effective
  \se\ varies strongly for very short times, but it tends quite rapidly, although
  in an oscillatory manner, to the constant value of the complex quasi-particle energy.

 \section{Illuminated alloy}\label{sec:resp_calc}

The electron response to an incident light pulse is conveniently expressed in terms
of the \NGF. The full formalism of the NGF is given in references
\cite{book:Haug} or \cite{book:Bonitz}. Here, only
the directly important relations will be given. Our approach is
characterized by two features.

(i)~We employ the NGF in the real time domain
in the Langreth--Wilkins representation \cite{LW}, that is we work with
$\Gr,\,\Gl,\,\Ga$. We prefer the LW triplet
over the usual Kadanoff--Baym pair $\Gl,\,\Gg$,
because in the present
case of a purely elastic scattering the  dynamics of the
propagators $\Gr,\,\Ga$ becomes decoupled from the particle
correlation function $\Gl$, and is much easier for numerical
solution. The LW equations also have a slightly more "physical"
appearance, suitable for interpretation.

(ii)~For non-interacting electrons,  time evolution is strictly unitary even
in a disordered system,
while the genuine field theoretic description relates to {\em configuration
averages}\cite{book:Gonis}, \cite{book:Sheng}. The \NGF\ needed here
can  be expressed in terms of configuration averages in a direct fashion following
\cite{we:zschrI}.

 Below, we describe the LW formalism for the disordered alloys and the
 computational procedures
 for the  electron propagator reacting to the light pulse, and for the particle
 correlation function.

 \subsection{Non-equilibrium \GF s}\label{sec:ngf}

 \subsubsection{General relations and LW conventions}
 \label{subsec:gen_rel}

   We work in the real time domain and
  the NGF is given by a 2$\times$2 matrix $\|G_{\alpha,\beta}\|$, which we choose
  to be in the Langreth--Wilkins representation \cite{LW}:
 \AS{1.2}
 \be
 \Gm\,=\,\left\|\ba{cc}
 G^{R}&G^{<}\\0&G^{A}\ea\right\|  \label{eq:LW}
 \ee
 Here, the component \GF s are one particle operators depending on two
 time arguments and satisfying the general relations
 \bea
 \Gr (t,t')&=&\{\Ga\}^{\dagger}(t',t) \label{eq:ra_symm}\\
 \Gl (t,t')&=&-\{\Gl\}^{\dagger}(t',t) \label{eq:gl_symm}\\
 \Gg (t,t')&=&-\{\Gg\}^{\dagger}(t',t) \label{eq:gg_symm}\\
 \Gr-\Ga&=& \Gg-\Gl \label{eq:sp_id}
 \eea
 The symmetry relation \Eq{eq:ra_symm} reduces the number of independent \GF s to two,
 while the last relation, \EEq{eq:sp_id} ({\em spectral identity}), links the LW
  representation with the usual Kadanoff--Baym pair $\Gl,\,\Gg$.

  The so-called LW algebra
 for the individual components of \Eq{eq:LW} follows easily from
 the usual rules for matrix operations, For example, introducing
 also the \se\ matrix $\Sim$, we have
 \be
 \Sim\,\Gm\,=\,\left\|\ba{cc}
 \Sir&\Sil\\0&\Sia\ea\right\|\cdot
 \left\|\ba{cc}
 G^{R}&G^{<}\\0&G^{A}\ea\right\|  \label{eq:LW_rules}
 \ee
 Thus,
 \bea
  \{\Sim\,\Gm\}^R&=&\Sir\,\Gr \label{eq:LW_ret}\\
  \{\Sim\,\Gm\}^<&=&\Sir\,\Gl+\Sil\,\Ga \label{eq:LW_less}
 \eea
 We use the convention that all quantities with no time variables
 shown are {\em double time}. Thus,
 $\Sim\rightleftharpoons\Si_{\alpha,\beta}\equiv\Si^X(t,t')$,
 $H_\SSC{mf}\rightleftharpoons
 H_\SSC{mf}(t)\,\delta(t-t')$, etc.
  Multiplications mean an operator multiplication and a time
  integral, like
  $\Sir\,\Gr\rightleftharpoons\int\,{\rm d}\bar{t}\,\Sir(t,\bar{t})\,\Gr(\bar{t},t')$.

 \subsubsection{Single alloy configuration}
 \label{subsec:single_conf}

  For non-interacting electrons,  time evolution is strictly unitary even
  in a disordered system:
  Each alloy configuration gives rise to a random Hermitian
  one-electron Hamiltonian
  ${\cal H}(t)$,
  which depends on time due to the
  effect of external fields (light pulse).
   The  non-random configuration
 independent or averaged quantities are denoted by italics, while the
 configuration  dependent quantities are denoted by
 rounded characters.

  For one configuration, we introduce the usual evolution operator ${\cal S}(t,t')$
  corresponding to ${\cal H}(t)$, and the one electron density matrix $\varrho(t)$.
  Selecting the earliest admissible
  time $t_0$, we introduce $\rho(t_0)$ as the initial condition
  for the NGF. Then, for all times $t,\,t'\ge t_0$, the random single
  configuration NGF is
  \AS{2.2}\be
  \Bfc{G}(t,t')=\left\|\ba{cc}
  {{\displaystyle 1} \over {\displaystyle\ci\hbar}}\,{\cal S}(t,t')\vartheta(t-t')&
  -\,{{\displaystyle 1} \over {\displaystyle\ci\hbar}}\,
  {\cal S}(t,t_0)\,\varrho(t_0)\,{\cal S}(t_0,t')\\
  0&-\,{{\displaystyle 1} \over {\displaystyle\ci\hbar}}\,{\cal S}(t,t')\vartheta(t'-t)\ea\right\|
  \label{eq:def_gcal}
  \ee

    Two properties of $\Bfc{G}$ are very important: First, the propagators
  ${\cal G}^R,\,{\cal G}^A$ do not
  depend on the particle distribution, and, for propagation from $t'$ to
  $t$, they only depend on ${\cal H}(\bar t)$ with $\bar t$ between
  the two times. Second, the correlation function $\Gcl$ does  in fact
  not depend on  the choice of the "initial time" $t_0$, if the relation
  $\varrho(t)={\cal S}(t,t_0)\,\varrho(t_0)\,{\cal S}(t_0,t)$ is employed.

 \subsubsection{Configuration average of \protect$\Bfc{G}$}
 \label{subsec:conf_ave}
  Observable quantities are presumably given by the
  configuration average of the particle distribution function
  $\rho = \ave{\varrho}$( see Sec. \ref{sec:physical_prop}, however); to obtain this quantity,
  we introduce the {\em
  averaged \NGF}\ starting from the \EEq{eq:def_gcal}:
  \be
  \Gm\,=\,\Ave{\Bfc{G}}
  \label{eq:ave_g}
  \ee
  These {\em configuration
  averages} ought to be treated by field theoretical means\cite{book:Gonis}, \cite{book:Sheng},
  \cite{we:zschrI}.

  The two propagator components are simple, for $X=R,\,A$,
  \be
  G^X(t,t')\,=\,\Ave{{\cal G}^X(t,t')}=\pm{ 1 \over {\ci\hbar}}\Ave{{\cal S}(t,t')}
  \vartheta(\pm[t-t'])
  \label{eq:ave_gra}
  \ee
  Now, we  obtain a
  semi--explicit expression for

  \bea
  \Gl&=&\Ave{{\cal G}^<}\,=\,\ci\hbar\Ave{{\cal G}^R\roo{\cal G}^A}
  \qquad (t\geq t_0,\ t'\geq t_0) \label{eq:g_less_def}
  \eea
  Here enters the \IC\ at $t=t_0$; the
  initial one-electron distribution is represented in the double time form as
  \bea
  \roo&=&\rho(t_0)\delta(t-t_0)\delta(t'-t_0)
  \label{eq:init_rho}
  \eea
  This \IC\ can be random, that is configuration dependent. We will
  exclude this possibility in our model by hypothesis. In this paper, we consider
  the case that {\em the
  initial distribution $\rho(t_0)$ is non-random}. By this, we eliminate the problem
  of initial correlations \cite{bonitz1999}.
   For the non-random
  initial condition, the averaged product in \Eq{eq:g_less_def} can be written as
  \AS{1.5}\bea
  \Gl&=&\underbrace{\ci\hbar{G}^R{\roo}{G}^A}_{\displaystyle {\sc coherent}}+
  \underbrace{{G}^R\Si^<{G}^A}_{\displaystyle {\sc incoherent}}
  \label{eq:g_less_dyson}
  \eea
  This equation {\em defines} $\Si^<$,  which is a regular two-time kernel in this case.
  The \EEq{eq:g_less_dyson} is then nothing else than the $<$-component of the \DE\
  for the \NGF\ matrix
  \Eq{eq:LW}. It is a good example of the suggestive LW form of the theory. It has
  the full $R\leftrightarrow A$ symmetry and a causal structure. The initial
  state evolves coherently in resemblance to unaveraged ${\cal G}^<$ (cf.
  \EEq{eq:g_less_def}), but it is evanescent. The population is replenished by an
  incoherent backscattering due to $\Si^<$.
 \EEq{eq:g_less_dyson}
 is also a starting point for developing equations of the Bethe--Salpeter type
 and the quantum transport equations.

 Given these results, we
 introduce the corresponding Dyson equation in the differential form
 by starting from the
 equations for the unaveraged \GF\ and performing the configuration average
 term by term.

  The  random Hamiltonian
  can be divided into its configuration average,
  the {\em mean field } Hamiltonian,
  and the fluctuation potential, that is the configuration dependent
  {\em random field}: ${\cal H}(t)=H_{\SSC{mf}}(t)+{\cal D}(t)$, as
  introduced by \EEq{eq:split_mf}.
 Both parts of ${\cal H}(t)$ might be time dependent at this level
 of generality, in contrast to the specific model, Eqs. \Eq{eq:hmf_expl},
 \Eq{eq:rand_fluct}  which
 we employ for explicit calculations in this
 paper.

  By  the Eqs. \Eq{eq:ave_gra}, \Eq{eq:g_less_def}, and \Eq{eq:split_mf},
 \bea
 (\Dt\,-\,H_\SSC{mf})\,\Bfc{G}\,-\,\Bfc{D}\,\Bfc{G}&=&\Bf{I}
 \label{eq:dyson_unav}\\
 (\Dt\,-\,H_\SSC{mf})\,\Gm\,-\,\Sim\,\Gm&=&\Bf{I}
 \label{eq:dyson_diff}
 \eea
  Here, $\partial_t\rightleftharpoons{{\partial} \over {\partial t}}\delta(t-t')$,
  $\Bf{I}\rightleftharpoons \delta_{\alpha,\beta} I
 \rightleftharpoons
 \delta_{\alpha,\beta}1_\SSC{op}\delta(t-t')$, and
 $\Bfc{D}\rightleftharpoons{\cal D}(t)\Bf{I}$.
 On going from \Eq{eq:dyson_unav} to
 \Eq{eq:dyson_diff}, we define  the \se\ matrix by setting
 $\Ave{\Bfc{D}\,\Bfc{G}}\equiv \Sim\,\Gm$.
 The possibility of such definition is itself another form of saying that
 the initial condition is non-random.
  A brief discussion of the random
  correlated initial condition is deferred to the Appendix \ref{app:corr_ic}.

 The  \DE\ \Eq{eq:dyson_diff}
 has to be supplemented by an actual form of the self-energy in order to
 become closed.
 In this paper, we will use the  \SCBA\ (SCBA), which  is a generic
 self-consistent conserving approximation appearing as a
 leading term either in the \SC\ multiple scattering expansion, or
 in the \SC\ perturbation expansion. The relationship between SCBA
 and the Coherent potential approximation within the multiple scattering
 context is discussed
 in \cite{we:zschrI}. Here, we use the \SC\ perturbation approach.
 This is based on the integral equations
 \bea
 \Bfc{G}&=&\Gm\,+\,\Gm\,(\Bfc{D}\,-\,\Sim)\,\Bfc{G}
 \label{eq:lw_eq_left}\\
 \Bfc{G}&=&\Gm\,+\,\Bfc{G}\,(\Bfc{D}\,-\,\Sim)\,\Gm
 \label{eq:lw_eq_right}
 \eea
 which follow from Eqs. \Eq{eq:dyson_unav}, \Eq{eq:dyson_diff}, and
 which can be iterated to yield a renormalized Born series in terms of the
 full propagators  $\Bfc{G}$ and the renormalized perturbation $\Bfc{D}\,-\,\Sim$.
 To obtain a closed equation for the \se, we
 iterate \Eq{eq:lw_eq_left} once using \Eq{eq:lw_eq_right}
 and configuration average
 respecting \EEq{eq:ave_d}:
 \bea
 \Sim&=&\Ave{\Bfc{D}\,\Bfc{G}\,\Bfc{D}}\,-\,\Sim\,\Gm\,\Sim
 \label{eq:sig_eq}
 \eea
 This yields an iterative expansion in terms of $\Bfc{D}$ and $\Bfc{G}$
 \AS{1.5}\be \ba{r@{}c@{}l}
 \Sim\,&=&\,\overbrace{\Ave{\Bfc{D}\Gm\Bfc{D}}}^{\displaystyle \Sim_\SSC{scba}}
 +\Ave{\Bfc{D}\Gm\Bfc{D}\Gm\Bfc{D}}+\\
     \ +\,&&\Ave{\Bfc{D}{\Gm}\Bfc{D}{\Gm}\Bfc{D}{\Gm}\Bfc{D}}
-\Ave{\Bfc{D}{\Gm}\Bfc{D}}{\Gm}\Ave{\Bfc{D}{\Gm}\Bfc{D}}
-\Ave{\Bfc{D}{\Gm}\Ave{\Bfc{D}{\Gm}\Bfc{D}}{\Gm}{\cal
D}}+\,\ldots
 \label{eq:matrix_scba}
 \ea\ee
The expansion starts from the second  order in ${\cal D}$, which
is our SCBA. In the fourth order, the expansion already deviates
from a simple geometrical series. In the language of diagrams, the
first subtraction excludes a reducible diagram, the second
suppresses double counting of self-energy insertions in the inner
propagator lines \cite{book:Gonis}.
 The physical approximation for $\Sim$, like a termination of the expansion at some
 order, is made
 consistently for all components of the self-energy, which guarantees
 its self-consistent conserving property. This is also the case of the SCBA.\\

 \subsection{Working form of the Dyson equations}
 \label{subsec:work_dyson}

 Now we write explicitly the equations for
 all components of the NGF in the form suited for numerical solution.
 These are obtained as the corresponding components of the matrix equations
 \Eq{eq:dyson_diff} and \Eq{eq:matrix_scba} with the use of the LW rules
  \Eq{eq:LW_rules} -- \Eq{eq:LW_less}. For each component, we have to specify:
  $\diamond$~~the equation of motion;{\ }
  $\diamond$~~\IC s;
  $\diamond$~~the specific form of the SCBA \se.

 We repeat that the equations will be written in the Galitskii picture with the
 convention that\\
 {\em the operators, Green functions, etc., will be written without
 tildes.} To return to the Schr\"{o}dinger picture, we will use the
 subscript $\ldots_{\bf S}$.

\subsubsection{Explicit equations for \protect \boldmath $\Gr$, $\Gl$}
\label{subsec:g_less_expl}

  For the propagators, we get
  \be
  (\Dt\,-\,H_\SSC{mf})\,{G}^{R,A}\,-\,\Si^{R,A}{G}^{R,A}=I
  \label{eq:ra_dyson_diff}
  \ee
 \be
    \Gr(t=t'+0,t')={\ci\hbar}^{-1}.1_\SSC{op}
 \label{eq:ra_dyson_ic}
 \ee
 \be
 \Si^{R,A}_\SSC{scba}=\Ave{{\cal D}{G}^{R,A}{\cal D}}
 \label{eq:ra_dyson_ba}
 \ee
In fact, only one of the propagators has to be found directly,
as the other one is given by \Eq{eq:ra_symm}.

 For the particle correlation function, the three relations are as follows\\ 
 $\diamond$~~ The equation of motion has no $\delta$-singularity: 
 \be 
 (\Dt\,-\,H_\SSC{mf}){G}^<\,=\,\Si^R\,{G}^<\,+\,\Si^<\,{G}^A\, 
 \label{eq:g_less_dyson_diff} 
 \ee 
$\diamond$~~ It has to be integrated for all times $t>t_0$ for any fixed
$t'\ge t_0$. Thus, the initial value is $G^<(t_0,t')$. For a non-random
initial distribution $\rho(t_0)$, we get immediately
\bea
G^<(t_0,t')&=&\rho(t_0)G^A(t_0,t')
\label{eq:g_less_ic}
\eea
$\diamond$~~ The SCBA self-energy $\Sil$ has a form analogous to the propagator
component:
\be
\Si^{<}_\SSC{scba}=\Ave{{\cal D}{G}^{<}{\cal D}}
\label{eq:g_less_ba}
\ee

We will return to the equations \Eq{eq:ra_dyson_diff} -- \Eq{eq:g_less_ba}
below and show the methods of their practical handling and the
computed electron dynamics. Before that, we will make our model fully explicit.

 First, we select the initial distribution function by assuming that the valence
 band  is completely occupied, the conduction band entirely empty:
 \be
 \rho(t_0)=\sum\limits_{{\rm k}\in\,{\sf BZ}}
 \ket{v{\rm k}}\bra{v{\rm k}} \equiv P_v
 \label{eq:rho_init}
 \ee
 As indicated, the $\rho(t_0)$ operator coincides with the projector on the
 valence band states, which is non-random and alloy composition independent by
 assumption, so that \Eq{eq:g_less_ic} applies.

 Second, all equations have to be specialized for the two band model of
 Sec.\ref{subsec:two_band}.
 Thus, all GF, etc., will be $k$-diagonal  2x2 matrices:
   $G^X\,\rightleftharpoons \| G^X_{ab}(k;t,t')\|,\
   \{X= R,A,<;\ a,b=c,v;\ k\in {\sf BZ}\}$
   and similarly for the self-energy. Disorder  is acting only in the conduction
   band, so that only the $cc$ components of the \se\ are non-zero.
   The self-energy becomes $k$-independent
   in the single-site
   approximation.
   By this, the general relation \Eq{eq:ra_dyson_ba}
   simplifies to  a scalar equation in the spirit of the dark
   relations \Eq{eq:scba_dark}:
   \be \ba{rcl}
   \Si^X_{cc}(t,t')&=&\gamma\cdot F^X_{cc}(t,t')\\
    F^X_{cc}(t,t')&=&N^{-1}\sum\limits_{k\in {\sf BZ}}G^X_{cc}(k;t,t')\\
   \gamma&=&c(1-c)\delta^2
   \label{eq:scba_explicit} \ea\ee

  \subsubsection{Separation of the dark \se}
 \label{subsec:dark_sep}

In principle, we have to solve the Dyson equations  \Eq{eq:ra_dyson_diff}
 for $\Gr$ and \Eq{eq:g_less_dyson_diff} for $\Gl$, and to  simultaneously obtain the
 self-energies from the self-consistent relations
 \Eq{eq:ra_dyson_ba}, \Eq{eq:g_less_ba}. These equations will first be modified,
 however, by singling out the dark constituent of the self-energy. For these
 modified equations, a self-consistent cycle leads to their stable solution.
 The self-energy will be divided into its dark
 and  induced components not only for computational, but also for physical reasons.
 The dark component analyzed in the preceding section describes the polaron
 effects including the short time dynamics and the long time quasi-particle
 behavior. The induced component
 reflects the coherent coupling between the
 excitation and the scattering. Proper inclusion of this coherence effect
 appears to be essential to keep the theory conserving and consistent, as is
 documented below.

 We  will separate out the {\em dark} part of the retarded \se; what remains, is
 the {\em induced} part:
     \be
     \Sir=\Sir_\SSC{dark}+\Sir_\SSC{induced}
     \label{eq:sig_split}
     \ee
      No such separation is needed for the particle component, because {\em all} of
     it is induced in our case:
     \be
     \Sil=\Sil_\SSC{induced}
     \label{eq:sig_less_split}
     \ee
     For convenience, the subscripts will be shortened to {\sc d} and {\sc i}.

     The equations \Eq{eq:ra_dyson_diff} and \Eq{eq:ra_dyson_ba}
     for the retarded propagator
      become:
     \be \ba{rcl}
     \Dt\,{G}^R&-&\overbrace{(H_\SSC{mf}\,+\,\Sir_\SSC{d})}^
     {\displaystyle H_\SSC{qp}}
     \,{G}^R\ =\ \Sir_\SSC{i}\,{G}^R,\quad
     t>t'\\
     \Sir_\SSC{i}&=&\Ave{{\cal D}({G}^R\,-\,{G}^R_\SSC{d}){\cal D}}
      \label{eq:ret_eq_def}
     \ea\ee
      The dark self-energy is known once for ever for
     a given system, and it appears as a permanent renormalisation ("polaron
     shift") at the l.h.s. of the Dyson equation. There is no coupling between this
     renormalizaton and the light. Without $\Sir_\SSC{i}$, the pulse would act as if it
     were exciting
     preexisting dark quasi-particles. Furthermore,
     the dark \se\ acts on the propagator
     almost as if it were local in time. For all these reasons, we denote
     the operator on the lhs of the Dyson equation by $H_\SSC{qp}$\ \ldots\ {\em
     an effective quasi-particle Hamiltonian}.

     Only the Dyson equation \Eq{eq:g_less_dyson_diff} is modified for $\Gl$ in view
     of Eq. \Eq{eq:sig_less_split},
      \be \ba{rcl}
    \Dt\,{G}^<&-&\overbrace{(H_\SSC{mf}\,+\,\Sir_\SSC{d})}^
    {\displaystyle H_\SSC{qp}}
    \,{G}^<\ =\
    \Sir_\SSC{ind}\,{G}^<
    \,+\,\Si^<\,{G}^A\,\quad
    t>t_0
    \label{eq:less_eq_def}
    \ea\ee
    while the equations \Eq{eq:g_less_ba}, \Eq{eq:g_less_ic} remain without
    change.
    This triplet \DE --SCBA--\IC\ is similar in many respects to the
    system for the retarded component. In particular, the lhs
    of the \DE\ has again the "polaron" interpretation of dark quasiparticles
    excited by the light pulse.

  \subsubsection{Induced part of the retarded \se}
 \label{subsec:ind_se}

     To see in better detail the
     structure of the induced part of the \se, we rewrite it explicitly:
     \be
     \Sir_{cc,\SSC{i}}(t,t')=\gamma N^{-1}\sum\limits_{{\rm k}\in{\sf BZ}}
     \left({G}^R_{cc}( {\rm k};t,t')\,-\,
     {G}^R_{cc,\SSC{d}}({\rm k};t,t')\right)
     \label{eq:siri_expl_sum}
     \ee
     Three points are apparent.

     First, $\Sir_\SSC{i}(t,t')=0$, if the interval
     $[t,t']$ has no overlap with the time span of the pulse, because then
     $\Gr = \Gr_\SSC{d}$. Thus  $\Sir_\SSC{i}$ vanishes in the quiescent time
     domain, and need not be computed for these times. This SCBA result
     coincides with its exact counterpart. Namely, in an exact treatment, the
     \GF\ is essentially a configuration average of a unitary evolution
     operator. The evolution is governed by a time local Hamiltonian, and the
     configuration average is also instantaneous. As a result, the averaged
     propagation
     from $t'$ to $t$ cannot depend on external fields acting outside this
     interval, just like in the approximate treatment.

     Second, the excitation by a pulse is not homogeneous in the whole Brillouin
     zone, acting predominantly in a strip of a width $\sim Q$ around the resonant
     energy. In the parts of the BZ far from this strip, the excitation is weak,
     and the full \GF\ deviates but weakly from the dark one. This averts the
     necessity to integrate over the whole zone -- in contrast to the dark \se\
     itself. Thus, only the physically relevant $k-$vectors are involved, and
     the details of the distant parts of the band structure have only a minor
     importance.

     This, in turn, reduces further the time domain, where the induced
     \se\ deviates from zero. While it vanishes exactly for  initial times
     $t'$ after the pulse, we may conclude now that it is also negligible for
     $t'$ well {\em before} the pulse. Namely, both $\Gr$ and $ \Gr_\SSC{d}$ are
     exponentially damped at a rate roughly corresponding to the resonant
     energy for all relevant ${\rm k}$; the whole induced \se\ is thus
     exponentially small in $t-t'$ as $t$ reaches the time of the pulse.

     Third, the induced part $\Sir_\SSC{i}$ is continuous for equal times,
     $t=t'$. Each of the self-energies has a $\gamma/\ci\hbar$ jump at equal
     times, and these two jumps exactly compensate. By extending the moment
     analysis of Sec. \ref{subsec:gf_dark_time} also to the full $\Gr$, it is
     possible to improve the short time estimate of $\Sir_\SSC{i}$. The moments
     depend on the inital time $t'$, and we quote without details that
     \EEq{eq:short_time} is generalized to
     \AS{1.9}
     \be \ba{rcll}
     \Gr_{cc}({\rm k};t,t')&\approx&
     {{\textstyle 1} \over {\textstyle \ci\hbar}}{\rm e}^
     {-{\scriptstyle \ci \over \scriptstyle \hbar}{\scriptstyle \ep_{\sc
     mf}({\rm k})\,(t-t')}}\times&\\
     &&\left( 1-(\gamma+Q^2\phi^2(t'))
     {{\textstyle (t-t')^2} \over {\textstyle \hbar^2}}
     - \ldots\right)&\ \\
     &\approx&
     {{\textstyle 1} \over {\textstyle \ci\hbar}}{\rm e}^
     {-{\scriptstyle \ci \over \scriptstyle \hbar}{\scriptstyle \ep_{\sc
     mf}({\rm k})\,(t-t')}}\times
     {\rm e}^{-(\gamma+Q^2\phi^2(t')){{\scriptstyle(t-t')^2}
     \over {\scriptstyle \hbar^2}}}
      &
     \label{eq:gf_full_short_time}
     \ea \ee
     For the shortest times, the contributions to the "action"
     due to light and to the disorder are
     independent and additive.
     Introducing this expansion into the \EEq{eq:siri_expl_sum}, we deduce the short
     time behavior of $\Sir_\SSC{i}$ in the form
     \be
     \Sir_{cc,\SSC{i}}(t,t')=\gamma Q^2\phi^2(t')){\rm e}^
     {-{\scriptstyle \ci \over \scriptstyle \hbar}{\scriptstyle \ave{\ep}\,(t-t')}}\,F_o(t-t')
     {{\textstyle (t-t')^2} \over {\textstyle \hbar^2}}
     +{\cal O}((t-t')^3)
     \label{eq:siri_short_time}
     \ee
     Here, $F_o(t)$ is the Fourier transform of the pure crystal local \GF,
     see \Eq{eq:reduction}. This is an important result, because it shows that
     the complicated short time behavior of the dark Green function does not
     enter the induced part of the \se, which gives it a welcome robustness with
     respect to approximations for \GF s.

     \subsubsection{Dark quasi-particles}
     \label{subsec:dark_qp}

     So far, the definition $H_\SSC{mf}\,+\,\Sir_\SSC{d}\,=\,H_\SSC{qp}$
     had only a symbolic meaning, and was not associated with an explicit
     introduction of the quasi-particle picture into the computations. There are
     reasons, both practical and theoretical, to investigate this
     possibility in more detail. On the practical side, the dark self-energy
     varies strongly in a very short time interval. To include it properly
     means to perform the integrations on the lhs of the \DE\ with a great care.
     At the same time, it is clear that beyond a short formation time, the dark
     propagators will assume the \WW\ form of a renormalized time exponential.
     Instead of directly using these WW propagators as a basis, we recall their
     phase form discussed in Sec. \ref{subsec:gf_dark_time}. The GF  of that
     section coincides with $G^R_{cc}$, the time $t$ is replaced by $t-t'$ in the
     non-stationary situation in the \EEq{eq:def_sigma}. For large time differences,
     the action of
     $\Sir_\SSC{d}$ will approach the WW limit as given by \Eq{eq:disp_law},
     $\sigma_k(t)\longrightarrow \sigma_k(\infty)= \gamma\cdot F_o(\ep_k\pm\ci 0)$
     regardless of the presence of the
     external field. On the other hand, for short times $t-t'$, we have
     shown in \EEq{eq:gf_full_short_time} that the phase variations caused by
     disorder and by the light are
     additive. Thus, the effect of the disorder
     and of the external disturbance appear as additive both in the short time
     and in the long time limits. Interpolating this behavior, we arrive at the
     following {\em quasi-particle approximation}:
     \AS{1.4}
     \be
     \left.H_\SSC{qp}\right|_k\,=\,\left\|\ba{cc}
     {\ep}_c(k)+\ave{\ep}+\sigma_k(t-t')&-Q\Phi(t)\\-Q\Phi(t)&{\ep}_v({k})+\hbar\Om\ea\right\|
     \label{eq:WW} \ee
     It is understood that this effective Hamiltonian acts on a Green function
     with time arguments $t,t'$ and $\sigma_k(t-t')$ acts as a
     {\em multiplicative}
     quantity. If the renormalization effects are not strong (slow energy
     variation of $F_0$), a further approximation, pure WW $\sigma_k(t-t')\approx
     \sigma_k(\infty)$, may be tested, as we did in \cite{we:noeks5}.

 \subsection{Computational schemes}
 \label{subsec:comp_scheme}

 \subsubsection{Computational scheme for the propagator}
 \label{subsec:comp_scheme_prop}

     We recall once more the important point that the propagator dynamics is
     fully independent of the particle distribution in our model with  elastic
     scattering. The propagators and the related self-energies can thus
     be computed from \EEq{eq:ret_eq_def} in an independent round of the whole calculation. This is a
     comparatively less demanding computational task. Only the retarded GF and \se\
     need to be obtained in a self-consistent cycle. The results can be stored, the advanced
     counterparts are obtained using the symmetry \Eq{eq:ra_symm}: $\Ga(t,t')=[\Gr(t',t)]^{\dag}$,
     $\Sia(t,t')=[\Sir(t',t)]^{\dag}$.
     In Fig. \ref{fig:flwch1}, we show the flowchart for  computing $\Sir_\SSC {I}$.
     For solving the Dyson equation, we used a fourth order adaptive Runge-Kutta-Fehlenberg
     solver. Interesting is  the seemingly counterintuitive  order of time loops.
     This is dictated by the fact that we are bound to make
     one step in the floating time, $t\longrightarrow t+\Delta t$, for all
     intermediate
     initial times $t>\bar{t}>t'$, so as to be able to calculate the product
     $\Sir_\SSC{i}\,{G}^R\longrightarrow\int{\rm d}\bar{t}
     \Sir_\SSC{i}(t+\Delta t,\bar{t})
     \,{G}^R(\bar{t},t')$ for the new time $t+\Delta t$.

 \subsubsection{Computational scheme for the particle function}
 \label{subsec:comp_scheme_part}
   On the whole, the Eqs. \Eq{eq:ret_eq_def} and \Eq{eq:less_eq_def} have an
   analogous structure.
   There are also marked differences.
  For $\Gl$,  the integration of the \DE\ starts at $t=t_0$ for any
  fixed $t'$. The advanced propagator enters \Eq{eq:less_eq_def}\ at two
  places, in the \IC\ and on the rhs of the \DE\ in the  integral
  $\int{\rm d}\bar{t}\Si^<(t,\bar{t})\,{G}^A(\bar{t},t')$. Thus,
  $G^A(\bar{t},t')$ is needed for $t_0\le\bar{t}\le t')$. It is readily available
  from the retarded propagator $G^R(t',\bar{t})$ calculated and stored
  beforehand, at the retarded stage, by means of the
  crossing symmetry \Eq{eq:ra_symm}. Another difference is that the
  self-consistent equations for $G^<$ and $\Si^<$ are
   linear, once the propagators are known, so that no iteration for obtaining
       $\Si^<$ is needed in principle, and the integration process runs only
  once.  An iteration of the process appears as necessary on the practical
  level, however. Namely, in contrast to the retarded case, where
  $\Si^R(t,t)=0$ is known, there is no universal value of $\Si^<(t,t)$, and this
  quantity must be estimated by extrapolation and refined by  iteration.
  The whole process is outlined in a flow chart shown in Fig.
 \ref{fig:flwch2}. For the particular Hamiltonian \Eq{eq:transf_ham} and
 ${\cal D}$ given by \Eq{eq:rand_fluct},
 the explicit form of the integral for $\Si^<$ is
\be
 \Si^<_{cc}(t,t')=\gamma N^{-1}\sum\limits_{{\rm k}\in{\sf BZ}}
 {G}^<_{cc}( {\rm k};t,t')
\label{eq:sili_expl_sum} \ee $$
\gamma=c(1-c)\delta^2$$ The number of primitive cells is denoted by $N$, the
Nordheim parameter $\gamma$ as the second cumulant of ${\cal D}$ measures the
disorder scattering strength in the SCBA.  No subtraction of the dark GF is
needed, because without illumination the electrons cannot leave the valence
band, $\Gl_\SSC{d}=P_v\Gl_\SSC{d}P_v$, while disorder is confined to the
conduction band.  All of the $cc$-component  of the full $\Gl$  is induced, as
stated in \EEq{eq:sig_less_split}. The integration in \Eq{eq:sili_expl_sum}
extends in fact only over a narrow shell around the one-photon resonance, just
like in the integral \Eq{eq:ret_eq_def} for $\Sir_\SSC{ind}$. There, however,
the subtraction is crucial.

    \section{ Numerical example}
     \label{sec:numerical}

     In this section, we show a  representative numerical example for the
     alloy model defined in Sections
     \ref{subsec:two_band} and \ref{subsec:parametrization}.
     The bare A-crystal band structure is characterized by the band
    gap $E_G = 1.5\,$eV, the electron and hole masses $m_c=0.4\,m_e$,
    $m_v=0.6\,m_e$. The bands are "proportional", with band widths 12$\,$eV,
    and 8$\,$eV, respectively. The band density of states is given in
    \ref{subsec:parametrization}. The alloy parameters are: $c = 0.05$, $\delta = -0.84\,$eV,
    so that $\ave{\ep}= 0.042\,$eV and $\gamma = 0.034\,\mbox{eV}^2$.
     We  use the basic frequency
     $\hbar\Omega=2\,$eV for a pure A-crystal
     and  adjust the frequency for the actual $c$ to 1.75$\,$eV according to
     Table \ref{tab:1}
     The exciting light pulse is specified in \ref{subsec:pulse_par} below.

     All parameters are
     selected so that they lead to pronounced but moderate
     effects. The formation times are shorter than either the
     pulse duration or the electron relaxation time, and the photoexcited electron
     density is sufficiently low that the $e-e$ relaxation time is longer than both
     these times.

     \subsection{Pulse parameters}
     \label{subsec:pulse_par}

     We have to specify the pulse strength $Q$ and its shape $\Phi(t)$,
     see \Eq{eq:interband_term}.
 The pulse strength is
     $Q =\frac{1}{2} e{\rm E}_\SSC{m}\bra{c{\rm k}\!\simeq\! 0}
 ({\rm e},{\rm r})\ket{v{\rm k}\!\simeq\! 0}$
 by \Eq{eq:interband_term}. The $x$-matrix element is
 estimated as the lattice spacing $a_0 \approx 10\,${\em a.u.}, corresponding to
 the momentum matrix element of about 0.5$\,${\em a.u.}
 As $\max\Phi=1$, we get
 the estimate ${\rm E}_\SSC{m}[{\rm Vm^{-1}}]=1.89\times 10^{9}\, Q[{\rm eV}]$
 for the peak value of the electric field.

 We use the "$\mbox{\rm sech}^2$" shape of the pulse and control its
  duration by a single parameter $t_\SSC{p}$:
 \be \ba{rcl}
 \Phi(t)&=&\varphi((t-t_{1/2})/t_\SSC{p})\\
 \varphi(t)&=&
 4/({\rm e}^{t}+{\rm e}^{-t})^2\ea
 \label{eq:def_phi}
 \ee
 The $t_{1/2}$ shift is defined by the condition
 $\Phi(0)=\half$. This shift and
 the full width at the half maximum of intensity
 are related to $t_\SSC{p}$ by\cite{const}
 \be \ba{rcl}
 t_{1/2}&=&t_\SSC{p}\cdot 0.88137\\
 t_\SSC{fwhm}&=&t_\SSC{p}\cdot 1.21169
 \label{eq:char_times}
 \ea \ee

 A convenient integral measure of the pulse
 is the resonant Rabi phase,
 \be
 \varphi_\SSC{r}=\int\limits_{-\infty}^{+\infty}{\rm d}t\,\Om_\SSC{r}(t)
 \label{eq:def_rph}
 \ee
 where  the Rabi frequency is introduced by
 \be
 \half\hbar\Om_\SSC{r}(t)=Q\phi (t)
 \label{eq:def_rf}
 \ee
 For our pulse shape \Eq{eq:def_phi} this gives
 \be \ba{rcl}
 \varphi_\SSC{r}&=&4{\hbar}^{-1}t_\SSC{p}Q\\
                 &=&3.3012{\hbar}^{-1}t_\SSC{fwhm}Q
 \label{eq:our_rph}
 \ea \ee
 We  use a pulse with  characteristics according to the Table \ref{tab:3}.

 \subsection{Propagators}
 \label{subsubsec:g_reta}

     \subsubsection{Calculated \protect\boldmath$\Sir_\SSC{I}(t,t^{\prime})$}
     \label{subsec:calc_sir}

     $\Gr$ and $\Sir$ are given by the Eqs. \Eq{eq:ret_eq_def} and
     \Eq{eq:siri_expl_sum}. These were solved as explained in \ref{subsec:comp_scheme}.
     First, we present in  Fig. \ref{fig:sirtt} the
     imaginary (dominant) part
     of the induced \se\ as a function of both times $t,\ t'$.
     This calculated shape manifests the properties predicted
     in Subsec. \ref{subsec:ind_se}.
     As a function of $t$ for a fixed initial time $t'$, it is strictly
     zero for $t<t'$ and it continuously assumes non-negative values on crossing
     the
     diagonal $t=t'$. For $t>t'$,
     it shows a marked
     memory effect, that is  time non-locality, or temporal coherence, and it ends
     with
     a  tail corresponding to the dark decay rate of  the  excitation after the end
     of the pulse.
     In the perpendicular direction, as a function of $t'$ for $t$ fixed,
     the $\Im\Sir\SSC{i}$
     profile is symmetric, decaying exponentially for times $t'$  well before
     the pulse arrival and ending continuously but abruptly at the upper limit.
       Across the time diagonal, the damping is much faster, as seen both in the
       main plot, and in the rotated insert. To motivate this, we note that in
       the Wigner coordinate $t-t'$, the exponential
      damping has a doubled decrement: $\Gr$'s in \Eq{eq:siri_expl_sum} depend only on
      $t - t'$ once both times are large enough, and for $t+t'= \mbox{const}$, we have
      $t - t'=2t-\mbox{const}$. Along the time diagonal, as a function
      of $t+t'$,  $\Im\Sir_\SSC{i}$ repeats well the shape of the pulse.

     The induced
     \se\ has a  much smaller magnitude  than  its dark counterpart, but it persists
     much longer. As a consequence, it is difficult to qualitatively guess the relative
     importance of both components of the \se\
     for the resulting propagators.

     \subsubsection{Calculated  \protect\boldmath$G^R(t,t^{\prime})$}
     \label{subsec:calc_gr}

     In the two-band model, the propagators are given by $2\times 2$ complex matrices
     $\|\Gr_{a,b}({\rm k};t,t')\|$.
     We show the behavior of the GF for just one point in the BZ,
     namely ${\rm k}\approx{\rm k}^{\times}$, by examining the time evolution of two
     quantities, $\Abs{\Gr_{cc}}^2$ and
     $\Abs{\Gr_{vv}}^2$. For a time development
     which started from a pure ${\rm k}^{\times}c$- or from a
     pure ${\rm k}^{\times}v$- state, respectively,
     they measure the probability that the particle stays in its initial state.
     In Fig. \ref{fig:gr_two}, we plot the two quantities as a function of $t$ and
     $t'$. For equal times, they start from a constant value $\hbar^{-2}$. For
     initial times well out of the pulse time interval, the decay of the $c$
     state resembles the dark alloy, while the $v$ state  does not decay at
     all. Within the pulse region, the very earliest evolution is still
     dominated by the dark \se. Soon, however, the external field controls the
     evolution, and the Rabi oscillation develops.

     This qualitative picture does not change, if the induced part of the \se\
     is switched off, as depicted by plotting two surfaces in each panel of the
     figure, which resemble each other quite closely. The differences are but
     quantitative. In other words, the dark quasi-particle approximation of Sec.
     \ref{subsec:dark_qp} neglecting $\Sir_\SSC{i}$ is qualitatively satisfactory.
     It has its serious problems, however, as we show now.

     \subsubsection{Testing the semi-group property}
     \label{subsec:test_semi}

     There exists a sensitive test of the quasi-particle behavior of the
     retarded \GF, which leads to an easy and convincing proof of the importance
     of the induced \se \cite{we:noeks5}. The true \WW\ dark quasi-particles, with
     zero illumination, would have an exponentially damped  modulus. A
     generalization proper for the dark quasi-particle approximation  in the WW limit,
     as defined
     in Sec. \ref{subsec:dark_qp}, employs the time-locality of the
     WW quasi-particle Hamiltonian to justify the {\em semi-group multiplicative
     property} of the propagator. This is summarized in the following equations:
      \AS{1.5}\be\ba{ccr@{\,\cdot\,}l}
   \ci\hbar {\cal G}^R(t,t')&=&\ci\hbar {\cal G}^R(t,t'')&
   \ci\hbar {\cal G}^R(t'',t')\\
     \ci\hbar {G}^R(t,t')&\neq&\ci\hbar {G}^R(t,t'')&\ci\hbar {G}^R(t'',t')\\
     \ci\hbar {G}^R_\SSC{ww}(t,t')&=&
     \ci\hbar {G}^R_\SSC{ww}(t,t'')&\ci\hbar {G}^R_\SSC{ww}(t'',t')\\
     &&\multicolumn{2}{c}{t\geq t''\geq t'}\\
     \label{eq:sg_prop}
     \ea\ee
     For the unaveraged GF, the equality follows from
     \EEq{eq:def_gcal}, while for the exact averaged GF, the {\em in}equality
     expresses the fact that a product of averages does not equal to the average
     of a product of two quantities. For the quasi-particle case, the equality is
     restored by the time locality of the effective Hamiltonian \Eq{eq:WW}.  We
     study again $\Gr$ for the resonant $k$-state starting this time at $t'=-0.2\,$ps.
     It is preferable to examine the squares of the full coherent amplitudes,
     $\Abs{\Gr_{cc}}^2+\Abs{\Gr_{vc}}^2$ and
     $\Abs{\Gr_{vv}}^2+\Abs{\Gr_{cv}}^2$.
     In Fig. \ref{fig:sg}, we see how the $c$-band state initially decays with
     the dark decay rate, but later  one Rabi flip is apparent. The $v$-band state
     is stationary at first. The
     pulse leads to its substantial depletion; a hint of the Rabi oscillation
     is also noticeable.
     The left panels incorporating $\Sir_\SSC{i}$ and the right panels using
     the  WW quasi-particle approximation lead to a picture  qualitatively
     resembling the exact result to a large degree.

     Now we start testing the
     semi-group property \Eq{eq:sg_prop}. As marked by squares on the profile of
     the pulse, several instants $t''$ were selected, at which we factorized the GF
     and plotted in various  thin lines the time evolution corresponding to
     $\ci\hbar {G}^R(t,t'')\ci\hbar {G}^R(t'',-0.2)$ for $t>t''$. While these
     thin lines coalesce with the full line of the non-factorized \GF\ in the
     quasi-particle case, the factorization for the true GF appears as approximately
     correct only
     for factorization times well before or well after the maximum of the pulse,
     when, of course, we deal with a nearly dark evolution, so that the time
     non-locality of the \se\ does not play an important role.
     We will see in Sec. \ref{subsubsec:total}, how these deviations from
     the quasi-particle behavior
     influence also the evolution of the photo-excited electron population.

     It should be pointed out that these results were obtained in the WW limit,
     because the discrepancy between both cases is particularly clear. Similar
     differences caused by the neglect of $\Sir_\SSC{i}$ appear also in the
     general case and are not peculiar to the WW approximation.

 \subsection{Particle correlation function}
 \label{subsec:g_less}

    \subsubsection{Calculated
    \protect\boldmath$\Sil(t,t^{\prime})$}
    \label{subsubsec:part_se}

    The 'less' component of the \se\ is the
    central quantity for a full
    description of the excitation and transport processes generated by the
    pulse: in the \BSE\ $\Sil$ appears as an irreducible vertex
    incorporating
    all correlations, which are in the present case due to the configuration
    average
    of the coherent multiple scattering of an {\em e-h} pair
    out of the equilibrium. As stressed already in Secs.
    \ref{subsec:dark_sep} and \ref{subsec:comp_scheme_part},  $\Sil$ would be zero without an
    excitation to the $c$ band
    which activates the disorder scattering.
     $\Sil$ has a coherent component resembling the pulse shape and
    width along the time diagonal $t - t' = 0$, as shown  in
    Fig. \ref{fig:sigma_less}. This  is similar to $\Sir_\SSC{ind}$,
    Fig. \ref{fig:sirtt}. A marked new feature, however, is the
    appearance of a tail of $\Sil$ which persists beyond the pulse duration. This is
    connected in a \SC\ manner with the photoexcited population in the $c$
    band. For the SCBA, we may see it clearly from the
    \EEq{eq:sili_expl_sum}, by which $\Sil$ is proportional to $\Tr
    G^<_{cc}$.

    Finally, we look in the direction
    of the time diagonal (the arrow in the figure) to inspect the profile
    of  $\Sil$ along  the other, $t+t'$, diagonal. In the small inserts,
    the plots of the real and the imaginary part of the \se\ are
    correspondingly  rotated and
     the  symmetry relations equivalent to
    \Eq{eq:gl_symm} are demonstrated. This represents a
    non-trivial check of the
    numerical approach used, because the integration procedure based
    on the differential \DE\ \Eq{eq:less_eq_def} is  not {\em a priori}
    symmetric with respect to both times.

    \subsubsection{Calculated
    \protect\boldmath$G^<(t,t^{\prime})$}
    \label{subsubsec:part_co}

    Fig. \ref{fig:g_less} shows
    just one representative
    $G^<_{cc}( {\rm k};t,t')$, namely that for ${\rm k}$ near the one-photon
    resonance. Symmetries \Eq{eq:gl_symm} are verified as a feature of
    individual terms of \Eq{eq:sili_expl_sum}, as shown in the format
    of  Fig.  \ref{fig:sigma_less}. We will
    concentrate on  the even part $\hbar\Im\,G^<_{cc}( {\rm k};t,t')$, because
    its time
    diagonal equals to the induced population of the $\ket{c{\rm k}}$ state.
    The induced transient
    is followed by pronounced structures slowly decaying along both time
    axes $t,\, t'$ and by a slow rise tending to saturation along the time
    diagonal $t=t'$. A closer look shows about one full
    wave along the diagonal, in agreement with the Rabi phase
    of our pulse being adjusted to  about $2\pi$. The side wings display
    only one rise followed by a decrease suggesting a phase variation $\approx\,\pi$.
    To understand the origin of this
    behavior better, we split in Fig. \ref{fig:g_less_coh} the \GF\ into
    its coherent and incoherent parts in accordance with \Eq{eq:g_less_dyson}.
    The side arms of $\Im\,G^<_{cc}$ appear to originate from its coherent
    part $\Im\,G^<_\SSC{coh}$. In fact, for one time fixed, the profile
    along the other time
    axis is proportional to $G^R$, so that its oscillation is given by the
    Rabi phase halved, and the attenuation at long times is
    $\hbar/\Im\,\Sir_\SSC{dark}$. Along the time diagonal,
    $G^<_\SSC{coh}$ is bilinear in  $G^R$, and the attenuation time
    equals now to the dark quasiparticle lifetime
    $\hbar/2\,\Im\,\Sir_\SSC{dark}$ leading to a much faster decay.
    The incoherent part  $\Im G^<_\SSC{incoh}$ appears with a time lag due to the
    multiple scattering delays. It is concentrated to a zone
    along the time diagonal, where it gradually supersedes the coherent part,
    so that it is responsible for the
    final rise of the full $\Im G^<$ in Fig. \ref{fig:g_less}.

    This description is somewhat
    oversimplified and  would be correct only in the dark quasiparticle
    approximation of \ref{subsec:dark_qp}, in which the exponentially
    decaying quasiparticles are driven by the incident light.
    In fact, the \EEq{eq:less_eq_def}  goes beyond the quasiparticle picture
    by incorporating  coherence between the light and the
    disorder as captured by $\Sir_\SSC{ind}$.
    This coherence was shown to be essential for a
    proper description of the propagators.
    It is equally important in the \EEq{eq:less_eq_def} for $\Gl$.
    We compare  for clarity the coherent parts $\Im\,G^<_\SSC{coh}$ with and
    without including $\Sir_\SSC{ind}$ in Fig. \ref{fig:g_less_noind}. While the
    two surfaces are superficially similar, they differ markedly at long times.
     It should be
    noted that while $\Sir_\SSC{ind}$ itself is a transient quantity,
    its coherent effects  persist.
    This is well illustrated in Fig. \ref{fig:g_less_diag}
    demonstrating the course of the full $\Im \Gl$ along the time diagonal.
    When \Eq{eq:less_eq_def} is solved without $\Sir_\SSC{ind}$, the
    asymptotic value of the state occupancy differs from the full
    solution.
    This has important implications for the particle number conservation,
    as will be discussed in detail in the next section.

    \section{Physical properties of the system}
 \label{sec:physical_prop}

  In general, observable physical properties are given by the equal-time limit
 of the particle correlation function\cite{book:Haug}. In the disorder case,
 this correspondence is not so simple in case the observable itself is
 represented by a random operator. For our model, however, the most important
 observables can be reduced to traces involving non-random operators.

 \subsection{One-electron distribution and observables}
 \label{subsec:rho_and_obs}

 For each configuration, the one-electron density matrix at a time $t$ is given
 by $\varrho(t)={\cal S}(t,t_0)\varrho(t_0){\cal S}(t_0,t)$. Thus, we have
 \bea
 \varrho(t)&=&-\ci\hbar{\cal G}^<(t,t)\\
 \rho(t)&\equiv&\ave{\varrho(t)}=-\ci\hbar{G}^<(t,t)
 \label{eq:rho_and_gless}
 \eea
 Consider now a  one-electron observable represented by an operator ${\cal X}$,
 which must be taken as configuration dependent in general. Its mean value
 susceptible to measurement is given by a double average,
 configuration and quantum statistical. It must in general be written as
 \be
 \dlangle{\cal X}\drangle=\Tr\Ave{{\cal X}\varrho}
 \label{eq:ave_xcal}
 \ee
 Thus, the configuration average concerns the product ${\cal X}\varrho$ and
 canot be reduced to the knowledge of the configuration averaged $\rho$. The
 problem is similar to that of the random initial condition, and a generalized
 procedure for \Eq{eq:ave_xcal} depending on the structure of ${\cal X}$ is
 required. This problem will not be treated presently, as we will limit
 ourselves to the {\em non-random} observables.

 \subsubsection{Averaged one-electron density matrix}
 \label{subsubsec:density_matrix}

 The configuration dependent equation of motion  for $\varrho$
 can be averaged with the result
  \bea
 {\partial \over {\partial t}}\varrho(t)\ -{1 \over {\ci\hbar}}
 [H_\SSC{mf}(t),\varrho(t)]\ &=& {1 \over {\ci\hbar}}[{\cal D}(t),\varrho(t)]
 \label{eq:rho_eom}\\
 {\partial \over {\partial t}}\Ave{\varrho(t)}-{1 \over {\ci\hbar}}
 \underbrace{[H_\SSC{mf}(t),\Ave{\varrho(t)}]}_{\displaystyle {\ci\hbar}\left.
 {\partial \over {\partial t}}\Ave{\varrho(t)}\right|_\SSC{drift}}
 &=&{1 \over {\ci\hbar}}\left\{\VV\right.
 \underbrace{\Si^RG^<-G^<\Si^A}_{\displaystyle {\ci\hbar}\left.
 {\partial \over {\partial t}}\Ave{\varrho(t)}\right|_\SSC{forw}}
 +\underbrace{\Si^<G^A-G^R\Si^<}_{\displaystyle {\ci\hbar}\left.
 {\partial \over {\partial t}}\Ave{\varrho(t)}\right|_\SSC{back}}
 \left.\VV\right\}_{t=t'}
 \label{eq:rho_ave_eom}
 \eea
 The r.h.s. of the \EEq{eq:rho_ave_eom} represents the collision rate
 $\left.{\partial \over {\partial t}}\Ave{\varrho(t)}\right|_\SSC{coll}$
  and the two terms
 correspond to the forward scattering and to the back scattering of the
 electrons. Altogether, a {\em precursor kinetic
 equation} is obtained, from which a true closed kinetic equation for $\rho$
 may be derived \cite{book:Haug}.
 This equation is  not independent of the the equation of
 motion for $G^<$ (and its conjugate), but it differs in the direction of integration
 in the $t,\,t'$
 plane: along the $t$ axis for \Eq{eq:g_less_dyson_diff}, but along
 the time diagonal $t=t'$ for \EEq{eq:rho_ave_eom}. In a consistent theory, the
 \GF s and the self-energies computed directly should turn \Eq{eq:rho_ave_eom}
 into a tautology. At this point, we may only state that our numerically obtained
 Green functions obey the precursor equation \Eq{eq:rho_ave_eom}, postponing
 all details to a publication about testing the ansatzes.

 From \Eq{eq:rho_eom}, we see explicitly that in  the exact theory  the particle
 number is conserved, that is $\Tr\Ave\varrho(t)$ is time  independent.
 A proof will now be given that the SCBA\ does the same,
 as expected from a
 conserving approximation, {\em cf.} Subsection \ref{subsec:conf_ave}.
 By the \EEq{eq:rho_ave_eom}, the conserving property
 is equivalent with the general criterion
 \be \ba{ccc}
 \Tr\left\{\VV\right.
 \Si^RG^<-G^<\Si^A+\Si^<G^A-G^R\Si^<\left.\VV\right\}_{t=t'}&=&0
 \label{eq:cons_crit}
 \ea\ee
 An explicit derivation of the conserving property both for the SCBA
 and for the CPA was previously obtained \cite{we:zschrI} in the special case of
 a rectangular light
 pulse. Here, we verify that SCBA satisfies the \EEq{eq:cons_crit} for an
 arbitrary pulse.
 The SCBA self-energy matrix is given by \Eq{eq:matrix_scba}.
 With the propagators written as $G^R=(G^>-G^<)\vartheta(t-t')$ and
 $G^A=(G^<-G^>)\vartheta(t'-t)$, the expression \Eq{eq:cons_crit}
 becomes
 \bea
 \left\langle\VV\right.\int\limits_{t_0}^{t}\!\!{\rm d}\bar{t}\;\Tr\left\{\VV\right.{\cal
 D}(t)\!\!&\!\!(\!\!&\!\!(G^>-G^<){\cal D}(\bar{t})G^<-G^<{\cal D}(\bar{t})(G^<-G^>)\\
      &\!\!+&\!\!G^<{\cal D}(\bar{t})(G^<-G^>)\,-\,(G^>-G^<){\cal D}(\bar{t})G^<\,)\left.\VV\right\}
      \left.\VV\right\rangle
 \label{eq:cons_test}
 \eea
 The order of the linear operations $\Ave{\ldots},\,\int$ and Tr has been
 interchanged and the cyclic property of trace employed. The \GF s to the left
 of ${\cal D}(\bar{t})$ have arguments $(t,\bar{t})$, those to the right have
 $(\bar{t},t)$. Clearly, the first term cancels with the fourth term, and the
 second  with the third. The result is zero, as required.

 \subsubsection{Observables and average values}
 \label{subsubsec:obs_and_ave}

 It will be
 convenient to compute the averages \Eq{eq:ave_xcal}  in the
 Schr\"{o}dinger picture; this amounts to the inverse transformation
 of the density matrix,
 \be\ba{rcl}
 \varrho_{\bf S}&=&O^{\dag}\varrho O\\
                &=&\varrho_{cc}+
                \varrho_{cv}{\rm e}^{-\ci\Omi t}
                +\varrho_{vc}{\rm e}^{\ci\Omi t}+
                \varrho_{vv}
 \label{eq:gali_to_schr}
 \ea\ee
 In the second line, the expression \Eq{eq:gali} for $O$ is used.

 The total particle
 number is a special case of the average value
 of an observable $X = 1_\SSC{op}$, which is non-random,
  so that the averaged $\rho=\ave{\varrho}$ was sufficient to use
  in the preceding subsection. There is a
 number of other important observables which are non-random in our model.
  In particular, the electric polarization vector equals to
 \bea
 {\bf P}&=&\Upsilon_0^{-1}\,N^{-1}\Tr(e{\bf r}\cdot\ave{\varrho_{\bf S}})
 \label{eq:ave_polaris}
 \eea
 because the position vector is represented by the non-random off-diagonal
 operators ${\bf r}\rightarrow {\bf r}_{cv}+{\bf r}_{vc} $.
 $\Upsilon_0$ denotes the volume of the primitive cell.
 The electric current density
 $\Bfc{J}=\Upsilon_0^{-1}\,N^{-1}e\dot{{\bf r}}=
 \Upsilon_0^{-1}\,N^{-1}(\ci\hbar)^{-1}[e{\bf r},{\cal H}]$,
 although a random operator, is also reduced to
 computing $\ave{\rho_{\bf S}}$, as
 \be\ba{rcl}
 \dlangle\Bfc{J}\drangle&=&\Upsilon_0^{-1}\,N^{-1}\Ave{
 \Tr((\ci\hbar)^{-1}[e{\bf r},{\cal H}]\varrho_{\bf S})}\\
  &=&\Upsilon_0^{-1}\,N^{-1}\Tr(e{\bf r}
 \underbrace{\Ave{(\ci\hbar)^{-1}[{\cal H},\varrho_{\bf S}]}}_
 {\textstyle \dot{\rho_{\bf S}}}
 )\\
 {\bf J}&=&{\partial \over {\partial t}}
 {\bf P}
 \label{eq:ave_current}
 \ea\ee

 The energy transfer between the external
 disturbance and the system can similarly be simplified in our model
 in which the optical field has no random component, that is the
 local field corrections are  negligible. In that case, the power
 absorbed per  second in a unit volume is
  \bea
  w&=&{\bf E}\cdot{\bf J}
  \label{eq:joule}
  \\
   &=&\Upsilon_0^{-1}\,N^{-1}
   \Tr(\overbrace{{\bf E}\cdot e{\bf r}}^{\textstyle -U(t)}\,
   \Ave{(\ci\hbar)^{-1}[{\cal H},\varrho_{\bf S}]}) \nnu\\
    &=&\Upsilon_0^{-1}\,N^{-1}
   \Tr(\Ave{(\ci\hbar)^{-1}[\underbrace{{\cal H}-U(t)}_
   {\textstyle \equiv{\cal H}_\SSC{dark}},{\cal H}]\,
   \varrho_{\bf S}})
  \label{eq:ave_power}
  \eea
  We introduce the Hamiltonian ${\cal H}_\SSC{dark}$ of electrons
  in the dark sample. The final form of the absorbed energy rate is then
  \bea
  w&=&\Upsilon_0^{-1}\,N^{-1}
  {\displaystyle {\partial \over {\partial t}}}
  \dlangle {\cal H}_\SSC{dark}
      \drangle
  \label{eq:ave_power_def}
  \eea
  The work done on the sample by the light turns into the change of the
  internal energy of the electrons, so that
   no true
 dissipation occurs, whether in the exact theory, or in the SCBA.
 This is to be expected in a theory without  sources of dissipation.
 Naturally, this sets an upper limit on the time interval in which such
 purely elastic theory may be valid.

     \subsection{Computed one-electron properties}
     \label{subsec:computed_rho}

     The computed double time \GF s were presented in Sec.
     \ref{subsec:g_less}.
     Here we continue with the results of numerical work
     considering the  time-diagonal
     $-\ci\hbar G^<(t,t)$,
     that is the one-electron density matrix $\rho$.

     \subsubsection{Time evolution of
     \protect\boldmath$\rho(t)$}
     \label{subsubsec:distr_f}

      First, we
     present in Fig. \ref{fig:rho_k} the full density matrix
     \Eq{eq:rho_and_gless} decomposed
     into its band- and $k$- vector dependent components. Because in our
     parabolic band regime the distribution is isotropic in the $k$ space,
     we use the bare electron energy instead of $k$. In the figure,
     $\Di={\ep}_c({\rm k})-{\ep}_c({\rm k}^\times)$,
     where ${\rm k}^\times$ corresponds to the weak field one photon
     resonance (see Fig. 2 and \EEq{eq:sc_crossing}).
     $\Di$ is
     related to the  usual detuning
     by a factor $m_v/(m_c+m_v)$. For the case presented, the
     spherical layer in the Brillouin zone
     corresponding  to Fig. \ref{fig:rho_k}
      is given by the inequalities
     $0.084\le\Abs{{\rm k}}\le 0.103\,${\em a.u.}, that is
     $0.134\le\Abs{{\rm k}}/k_X\le 0.170$. Here, $k_X=2\pi/a_0=0.605\,$
     {\em a.u.}.

     The diagonal panels show the excitation into the conduction band and
     the corresponding depletion of the valence band. The band diagonal
     elements are real and equal to the occupation numbers $n_{c,v}(\rm k)$.
        The overall effect corresponds to  the expectation.
     A coherent transient excites a broad region in the BZ as an energy
     uncertainty effect. It is followed by the
     persistent excited population in a narrow strip around the resonance,
     which is the strongest for small
     detunings.

     We use the off-diagonal
     panels to plot $\Re\,\rho_{cv}$ and $\Im\,\rho_{cv}$ which together
     give both off-diagonal elements $\rho_{cv}=\rho^*_{vc}$. This is
     the polarization part of the density matrix.
     It shows a complex oscillatory
     behavior with an approximate mirror symmetry around the zero
     detuning. The polarization is attenuated with the characteristic time
     $\hbar/\Im\,\Sir_\SSC{dark}({\rm k})$, that is twice slower than the
     diagonal transient. This can be understood from \Eq{eq:g_less_dyson}
     and \Eq{eq:rho_and_gless} by a reasoning similar to that in Sec.
     \ref{subsubsec:part_co}: the relevant quantities to be
     compared are $\Gr_{cv}(t,t_0)\varrho(t_0)\Ga_{vc}(t_0,t)$
     and, say,
     $\Gr_{cv}(t,t_0)\varrho(t_0)\Ga_{vv}(t_0,t)$. Both $G$
     factors decay in the former expression, only one in the latter
     one.
     We note that we intentionally do not introduce a semi-empirical
     dephasing time leaving the loss of coherence entirely to the effects
     of disorder.

     \subsubsection{Time evolution of
     observable properties}
     \label{subsubsec:total}

     Figure \ref{fig:distr_tot} presents, in a matrix
     arrangement, the total effect of the pulse integrated over the BZ,
     \be
     \rho_{ab}^\SSC{tot}(t)=N^{-1}\sum\limits_k\rho_{ab}({\rm k},t),
     \quad a,b=c,v
     \label{eq:rho_tot}
     \ee
     Full lines in all figures give the result of a direct GF
     computation.
     The diagonal panels show the total band populations per cell.
     An almost precise
     conservation law $\rho^\SSC{tot}_{cc}+\rho^\SSC{tot}_{vv}=1$
     is a satisfactory result, because the two contributions are computed
     independently on the basis of constituents which do not show much
     resemblance
     individually. The coherent parts do not compensate (thin line), and the deficit
     is compensated for by the backscattered flow.
     Further, we return for the last time to the importance
     of the induced part of the particle self-energy, $\Sir_\SSC{ind}$.
     By a dotted line, we plot the result, if this \se\ component is
     neglected and thus the dark quasiparticle approximation is used.
     Very clearly, the particle number conservation is strongly violated,
     and the error occurs in the valence band. This is in good
     correspondence with the tests of the semi-group property in Sec.
     \ref{subsec:test_semi}
     , where the most pronounced deviations also occurred
     for the valence band \GF. This is most likely due to the fact that
     our valence band is not affected by the disorder directly, and its
     behavior is dictated by the indirect influence of the conduction band
     in a rather sensitive manner.

     The off-diagonal component $\rho^\SSC{tot}_{vc}$ is shown
     in the $vc$ corner of Fig. \ref{fig:distr_tot} by components,
     in the $cv$ corner in the semi-logarithmic form:
     $\rho^\SSC{tot}_{vc}=R{e}^{\ci\varphi}$. All quantities shown
     correspond to the Galitskii envelope. They are related to the
     induced polarization per cell by
      \bea
 \Upsilon_0{\bf P}&=&e{\rm x}_{cv}\left\{\rho^\SSC{tot}_{vc}
 {e}^{-\ci\Omi t} + {\rm c.\,c.} \right\}\\
 &=&e{\rm x}_{cv}R(t)\cos(\Omi t+\varphi(t))
 \label{eq:polaris_schr}
 \eea

     Total polarization oscillates rapidly within the envelope $\pm R$.
     We compare its amplitude with the pulse envelope scaled to the
     rising part of polarization. The pulse and the response initially
     appear to
     be proportional  having  constant their ratio and phase difference.
     Around the pulse peak, the response is sub-linear, while the phase
     changes sign and rises close to $\half\pi$.
     Polarization vanishes right after the pulse: practically
     all of the polarization is associated with
     the coherent part of the excitation. The rapid attenuation of the
     polarization is no doubt caused by a cancellation due to
     dephasing of the rapid oscillations of $\rho_{vc}({\rm k}, t)$
     seen in the previous figure.

  Using Eqs. \Eq{eq:joule}, \Eq{eq:ave_current} and
  \Eq{eq:polaris_schr}, we get for the power absorbed per primitive cell
  and averaged over the $\Omi$  cycle
  \be\ba{r@{\,}c@{\,}l}
  \Upsilon_0\,w &=& 2Q\Phi(t)\left\{\Re({\textstyle{\partial \over {\partial t}}}
  \rho_{vc})
  -\Omi\,\Im\rho_{vc}\right\}\\
  &=& 2Q\Phi(t)\left\{{\textstyle{\partial \over {\partial t}}}
  (R\,\cos\varphi)
  -\Omi\,R\sin\,\varphi
  \right\}\ea
 \label{eq:power_schr}
 \ee
 The $\Omi$ oscillations  are rapid compared to variation of the
  Galitskii amplitudes (RWA condition) so that the second
 term in the braces is dominant. The phase angle between the
 current and the field is then basically $\varphi + \half\pi$.
 It evolves from a small positive value $\approx 0.095\pi$ to a
 saturation value close to $\pi$:
 the energy is absorbed at first, then partly returned
 to the field, as shown in Fig \ref{fig:joule}. The integral
 $W=\int^t\!{\rm d}t'\,w(t')$ multiplied by $\Upsilon_0$ is the energy
 transferred between the pulse and one primitive cell up to the time $t$.
 It is also shown in the figure. The ratio
 $\Upsilon_0 W(t)/\rho^\SSC{tot}_{cc}(t)$ for times after the pulse
 gives the energy absorbed per a photoexcited electron; it corresponds
 closely to the basic frequency of the pulse, $\Omi =1.75\,$eV.

 Considering the real magnitude of the photoexcitation, we have two electrons
 per primitive cell in the fully occupied valence band, that is the bulk
 electron density is $8/a_0^3= 3.3\times 10^{27}\,{\rm m}^{-3}$. With the
 fractional excitation $\sim 10^{-4}$, this yields
 $10^{23}\div 10^{24}\,{\rm m}^{-3}$ of the
 electrons excited to the conduction band. Such densities are low enough to
 justify that we neglect the $e-e$ collisions during the time interval
 considered, {\em cf.} \cite{book:Bonitz}. The induced polarization
 \Eq{eq:ave_polaris} can be estimated by
 ${\rm P}={8 / {a_0^3}}\,\rho_{cv}^\SSC{tot}\cdot ea_0$
 as compared with the vacuum electric induction
 ${\rm D}_\SSC{vac}=\varepsilon_0 {\rm E}_\SSC{m}$.
 Both quantities are comparable and have
 the $\sim  10^{-4}\,\mbox{Cm}^{-2}$ order of magnitude.

 \section{Conclusion}
 \label{conclusion}

 In this paper, we have addressed the problem of the response of electrons in
 semiconductors to a short light pulse. Of the three principal scattering
 mechanisms, we concentrated on the impurity scattering, and to make it
 pronounced, the case of concentrated semiconductor alloys was studied. For
 extremely short times, the disorder scattering in alloys may typically be
 dominant. The scattering regime is easily adjusted by varying the alloy
 composition.

 The physical picture obtained has the following features:\\ \DIA The alloy
 scattering in chemically saturated alloys is short range affecting thus large
 parts of the Brillouin zone.\\ \DIA It is elastic, which makes some coherence
 effects pronounced.\\ \DIA Finally, it also acts  in the dark, so that the dark
 polaron effect, {\em i.e.} use of dressed terminal states is essential.
 Typically, these dark states have a well-defined quasi-particle nature. \\ \DIA
 The illuminated state cannot be reduced to a redistribution of electrons among
 the dark states, and the induced part of the electron self-energy describes a
 coherent time non-local coupling between the alloy scattering and the
 excitation process.\\ \DIA As shown by a cumulant analysis, for very short
 times the dark component describing fast quasi-particle formation and the
 delayed induced component contribute to the self-energy additively, so that the
 formation process is insensitive to the illumination.\\ \DIA  The coherence between
 the light and scattering significantly affects the photoexcited distribution; in
 particular, it conditions the particle number and energy conservation.

 All these conclusions are obtained analytically and confirmed numerically.

 The paper discusses also technical questions. The NGF describing the process
 are obtained by a direct solution of equations equivalent to the Kadanoff-Baym
 equations. The asymmetric LW choice of computing first the propagators and
 using these as an input in the equation for the particle correlation function
 is advantageous in the present case of an elastic scattering. The
 isotropic scattering would require $k$-integration over the whole BZ. This is
 overcome by separating out the dark self-energy and integrating explicitly only
 the induced part, stemming from a narrow slice around the resonance in the
 $k$-space. To suppress rapid oscillations of the integrand, we work in the
 Galitskii representation, in which the pulse is represented by its envelope,
 while the high basic frequency enters only as a relative shift of the bands.

  For the future work, this paper suggests several directions as an immediate sequel:\\
  (i) It will be important to extend the NGF approach to the Coherent
  potential approximation, which is a self-consistent intermediate theory
  suitable for all realistic alloy scattering strengths.\\
  (ii) Then random ("correlated") initial conditions will play a non-trivial role,
   and should be incorporated.\\
  (iii) These analytical approximations for the NGF may be, in the alloy case,
  compared with a direct simulation for explicitly generated random alloy
  configurations and a numerically performed configuration average.

  Independently of these extensions and modifications, the results of the present work
  can readily be used for an analysis of various methods to develop quantum kinetic
  equations based on an ansatz generalizing the KBA, and for a comparison of the
  resulting approximate solutions with the NGF results.
  Such analysis should serve to find the validity limits of the individual
  ansatzes and the underlying physical explanation.
  A communication concerning these questions is also in preparation.

 \section{Acknowledgment}
 This work was supported by the Grant Agency of the Czech Republic under the
 project number 202/00/0643.

{\appendix
\section{\protect$\Gl$ with initial correlations}

\label{app:corr_ic}

Like in the main text, the basic definition of $\Gl$ is the
\EEq{eq:g_less_def}, but now with the initial condition \Eq{eq:init_rho}at the
initial moment $t_0$ being random. We express the propagators as ${\cal
G}^R=\Gr+\Gr({\cal D}-\Si^R){\cal G}^R$,
 ${\cal G}^A={G}^A+{\cal G}^A({\cal D}\,-\,\Si^A){G}^A$.
 The expresssion $\Gl=\ci\hbar\Ave{{\cal G}^R\roo{\cal G}^A}$ becomes
\bea \Gl   &=&\ci\hbar\left(\VV {G}^R\ave{\roo}{G}^A\right. \nnu\\
   &+&{G}^R\Ave{\roo{\cal G}^A({\cal D}\,-\,\Si^A)}{G}^A
   \,+\,{G}^R\Ave{({\cal D}\,-\,\Si^R){\cal G}^R\roo}{G}^A \nnu\\
   &+&\left.\VV{G}^R\Ave{({\cal D}\,-\,\Si^R){\cal G}^R\roo
   {\cal G}^A({\cal D}\,-\,\Si^A)}{G}^A
   \right)\\&&\qquad (t\geq t_0,\ t'\geq t_0)
 \label{eqap:g_less_expl}
\eea

 \n The three lines of \Eq{eqap:g_less_expl} correspond to an averaged
 ("uncorrelated") initial condition, to the random field analogue of the
 correlated \IC s, and to the genuine particle correlation, respectively. These
 terms have the same general structure $G^R\,\dots\,G^A$, but they differ in the
 number of inner times equal to $t_0$, two, one, and none. We make this explicit
 using the "$\VV_o$" subscripts and  define
 \AS{2.0}\bea \VV_{o\!}\Lambda^{<}_{\ }
 &\equiv&\ci\hbar\Ave{\roo{\cal G}^A({\cal D}\,-\,\Si^A)}
 \label{eqap:lambda_left}\\
 \VV_{\ }\Lambda^{<}_o
 &\equiv&\ci\hbar\Ave{({\cal D}\,-\,\Si^R){\cal G}^R\roo}
 \label{eqap:lambda_right}\\
 \Si^<&\equiv&\Ave{({\cal D}\,-\,\Si^R){\cal G}^<({\cal D}\,-\,\Si^A)}
 \label{eqap:sigma_less_exact} \eea
 With
 these definitions, the final form of $\Gl$ is
 \AS{1.5}\bea \Gl&=&\ci\hbar{G}^R
 \ave{\roo}{G}^A+ {G}^R\VV_{o\!}\Lambda^{<}{G}^A+
 {G}^R\Lambda^{<}_o{G}^A+{G}^R\Si^<{G}^A
 \label{eqap:g_less_dyson} \eea
 This equation
 extends the \EEq{eq:g_less_dyson} by incorporating a memory effect caused by
 the disorder induced initial correlations. In all other respects, comments made
 to \Eq{eq:g_less_dyson} apply here equally well.

 The diferential equation analogous to the \DE\ \Eq{eq:g_less_dyson_diff} is
 obtained, if we differentiate \Eq{eqap:g_less_dyson} from the left, and employ
 \Eq{eq:ra_dyson_diff} for $G^R$ in the form
 $$(\Dt-H_\SSC{mf}){G}^R\,=\,\Si^R\,{G}^R+I.$$ For $t>t_0$,
 the $\delta$-singularity is effective for only two terms of \Eq{eqap:g_less_dyson},
 and the resulting equation reads
 \bea
 (\Dt\,-\,H_\SSC{mf}){G}^<\,=\,\Si^R\,{G}^<\,+\,\Si^<\,{G}^A\,+\,\Lambda^{<}_o{G}^A
 \label{eqap:g_less_dyson_diff}
 \eea
 The two seemingly missing terms reappear in the initial condition:
 \bea
 G^<(t_0,t')&=&\left[ \ave{\roo}{G}^A+
 \VV_{o\!}\Lambda^{<}{G}^A\right]_{t=t_0,t'}
 \label{eqap:g_less_ic} \eea
 It remains to develop an approximation scheme for the self-energy,
 both for the regular part $\Sil$
 and for the singular part $\VV_{o\!}\Lambda^{<}+\Lambda^{<}_o$. The two
 $\Lambda\,'$s are equivalent:
 \bea
 \VV_{o\!}\Lambda^{<}(t,t')&=&-\left\{\Lambda^{<}_o(t',t)\right\}^{\dag}
 \label{eqap:lambda_symm} \eea
 so that the whole
 $<$ component of the \se\ obeys the symmetry \Eq{eq:gl_symm}. Considering, say,
 $\Lambda^{<}_o$, we see from \Eq{eqap:lambda_right} that it only involves
 retarded quantities, which do not depend on the initial conditions for elastic
 scattering. Thus, $\Lambda^{<}_o$ is a linear functional of the initial
 distribution $\rho_0$, and it does not involve any particle-hole correlation.
 In other words,  $\Lambda\,'$s can computed on the propagator level. The
 averaging procedure leading to \Eq{eqap:lambda_right} depends on the initial
 distribution, however.

 The  \EEq{eqap:sigma_less_exact} for $\Sil$ does not explicitly depend on the
 initial condition. In the SCBA, it will have the form
 $\Si^{<}_\SSC{scba}=\Ave{{\cal D}{G}^{<}{\cal D}}$ identical with
 \Eq{eq:g_less_ba}, but with $G^{<}$ given by \Eq{eqap:g_less_dyson} and
 \Eq{eqap:g_less_ic}. Only in higher order approximations, the genuine
 correlations of the three-terminal type, like
 $\Ave{{\cal D}\Gr(\roo-\ave{\roo})\Ga{\cal D}}$,
 start playing their role, and must be systematically accounted for.

 \section{Random observables}
 \label{app:rand_obs}

 Let us return to \Eq{eq:ave_xcal}, but write it in the Schr\"{o}dinger
 picture, {\em cf.} \Eq{eq:gali_to_schr}:
 $$
 \dlangle{\cal X}\drangle=\Tr\Ave{{\cal X}\varrho_{\bf S}}
 \eqno(\ref{eq:gali_to_schr}{\rm S})
 \label{eqap:ave_xcal}
  $$
  If the observable ${\cal X}$ is random itself, the double average has
  to be performed {\em at the end}, as discussed in the main text.
  The average can be given two different forms:
  \bea
  \dlangle{\cal X}\drangle_{t}&=&
  \hbar^2\Tr\Ave{{\cal X}{\cal G}^R(t,t_0)\roo{\cal G}^A(t_0,t)}
 \label{eqap:ave_symm}\\
                          &=&
    -{\ci\hbar}\Tr\Ave{{\cal X}{\cal G}^<(t,t'\rightarrow t)}
 \label{eqap:ave_gl}
 \eea
 The first form, \EEq{eqap:ave_symm}, makes explicit that the random
 initial condition and a random observable enter the averages in a
 symmetrical way, and any general technique for their evaluation has to
 be built up symmetrically. The second form of
 $\dlangle{\cal X}\drangle$, seemingly trivial, replaces the density
 matrix by the particle correlation function, and this may lead to
 explicit averaging procedures in special cases.

  As an important example, let us consider the dark Hamiltonian in our
  model and
  calculate $\dlangle {\cal H}_\SSC{dark} \drangle$, a quantity related
  to the Joule work by \Eq{eq:ave_power_def}.
  This does not go
  with $\ave{\rho}$, but we may proceed as follows:
  \be\ba{rcl}
  {\ci\hbar}{\partial \over {\partial t}}{\cal G}^<&=&
  ({\cal H}_\SSC{dark}+ U(t)){\cal G}^<\\
  \Ave{{\cal H}_\SSC{dark}{\cal G}^<}&=&
  {\ci\hbar}{\partial \over {\partial t}}G^<- U(t)G^<\\
  \dlangle{{\cal H}_\SSC{dark}}\drangle_t&=&
  \Tr\Ave{{\cal H}_\SSC{dark}(-{\ci\hbar}){\cal G}^<(t,t'\rightarrow t)}\\
  &=&{\hbar}^2\Tr({\partial \over {\partial t}}G^<(t,t'\rightarrow t))
  -\Tr( U(t){\rho})
   \label{eqap:hav_dir}
  \ea\ee
  Use is made of the fact that $U$ is non-random, and its average
  reduces to the use of $\rho=\ave{\varrho}$. The essential point is,
  however, that
  the derivative on the r.h.s. is NOT a derivative of the density matrix.
  The trick is similar to the computation of  correlation energy
  using the one-particle GF in usual many-body theory.
  The derivative  can be found directly (it would be enough to store it
  during
  the solution of the Dyson eq. for $G^<$), or it could be expressed
  using
  this DE (see \EEq{eq:g_less_dyson_diff}).

  Thus, we can  do two things: either get $\dlangle{{\cal
  H}_\SSC{dark}}\drangle_t$ directly from \Eq{eqap:hav_dir},
  or express it in terms of
  the integral of the "Joule heat", $W(t)=\int^t\!{\rm d}t'\,w(t')$ , using
  \Eq{eq:ave_power_def}. The two results should be the same. This
  provides another criterion for the conserving nature of any
  approximations involved in the calculation.

 \newpage

 \begin{table} 
 \caption{Concentration dependence of the alloy electron structure:
 weak field  one photon resonant energy , level broadening and particle lifetime}
 \begin{center}
 \begin{tabular}{cccc}
 $c$&$\hbar\Om^\times/\,$eV&$\Abs{\Im z_k}/\,$eV&$\tau/\,$fs\\ \hline
 0.00&1.80&0.0000&$\infty$\\ 0.05&1.75&0.0020&165\\ 0.15&1.63&0.0059&\ 53\\
 0.50&1.34&0.0105&\ 31\\
 \label{tab:1}\end{tabular}
 \end{center}
 \end{table}

 \begin{table} 
 \caption{Lowest moments and cumulants:  exact and  in the SCBA } \bec
 \AS{1.5}
 \begin{tabular}{cccccc}
 $\Bf{p}$&$\Bf{M_p}$ {\sc scba}&$\Bf{M_p}$ {\sc exact}&$\Bf{C_p}$ {\sc
 scba}&$\Bf{C_p}$ {\sc exact}\\ \hline $\VV$ 0&1&1&1&1\\ 1&0&0&0&0\\
 2&$\gamma$&$\gamma$&$\gamma$&$\gamma$\\ 3&$\gamma\mu_1$&
 $\gamma\mu_1+\gamma(c_B-c_A)\delta$&$\gamma\mu_1$&$\gamma\mu_1+\gamma(c_B-c_A)
 \delta$\\ 4&$\gamma\mu_2+{\bf 2}\gamma^2$&$
 \gamma\mu_2+2\gamma(c_B-c_A)\delta\mu_1+$&$\gamma\mu_2 {\bf -1}\gamma^2$&
 $\gamma\mu_2+2\gamma(c_B-c_A)\delta\mu_1+$\\
  &&$\qquad+\gamma(\delta^2-{\bf 3}\gamma)$&&$\qquad+\gamma(\delta^2-{\bf 6}\gamma)$\\
 \label{tab:2}\end{tabular} \enc
 \end{table}

  \begin{table} 
  \caption{Characteristic parameters  of the pulse}
  \bec
  \begin{tabular}{ccccc}
  $t_\SSC{p}[{\rm ps}]$&$Q[{\rm eV}]$&
  ${\rm E}_\SSC{m}[{\rm Vm^{-1}}]$&$t_\SSC{fwhm}[{\rm ps}]$
  &$\varphi_\SSC{r}$\\ \hline
  0.1&0.01&1.89${\times 10^7}$&0.121&1.93$\pi$\\
  \label{tab:3}\end{tabular}
  \enc
  \end{table}
\clearpage
 \begin{figure}
 \caption{Local \GF\ $F_o(E+\ci\,0)$ of the pure A crystal for energies between
 the band edges 1.5$\,$eV and 13.5$\,$eV  on the real axis. }
 \label{fig:f_o}
 \end{figure}

 \begin{figure}
 \caption{Renormalized dispersion law $z_k$ as a function of bare energy $\ep_k$
 for four values of concentration $c=0.;\,0.05;\,0.17;\,0.5$. Dotted lines:
 mean field rigid shift of the whole band. Thick line: full dispersion law with
 polaron shift included. Energy dependent band broadening
 shown by lining each of the quasi-particle energy by thin line at a distance
 $\pm \Im z_k$. Dot-dashed lines: shifted valence band ${\ep}_v({\rm k})
  +\hbar\Om^\times(c)$. } \label{fig:z_k}
 \end{figure}

 \begin{figure}
 \caption{Renormalization constant as a function of bare energy for the same
 values of concentration as in Fig.2. Upper part (thick lines):
  modulus $\Abs{Z_k} -1$. Lower part (thin lines):
 phase $\arg\,(Z_k)$.} \label{fig:Z_k}
 \end{figure}

 \begin{figure}
 \caption{Evolution of the "action" $\varsigma(t)$
  for dark propagator on femtosecond time scale. Fig. 4a: real part, Fig. 4b:
 imaginary part. Full line:  full SCBA solution.
 Dashed line:  pole approximation \Eq{eq:phase_pole} in time domain.
 Detail for the shortest times $\leq\,$0.3 fs (small windows) expanded in the insert
 in the left lower corner and compared with the lowest (full dots) and
 next lowest (open dots) order  cumulant expansions.
 The insert in the right upper corner:
 effective self-energy $\sigma_k$, \EEq{eq:sigma_varsigma} (full
 line) tends rapidly to the constant
 quasi-particle (pole) energy (dashed line).}
 \label{fig:varsigma}
 \end{figure}

 \begin{figure}
 \caption{Schematic flow chart  for calculation of $\Sigma_I^R(t,t')$.
 Input: dark QP for $p=0$. Iteration counter: $p$. Output: $\Sigma_I^R(t,t')$
 and $\Gr({\rm k};t,t')$ (not indicated).}
 \label{fig:flwch1}
 \end{figure}

\begin{figure}
\caption{Schematic flow-chart for calculation of $\Sil(t,t')$. Input:
propagator components $\Sir_\SSC{d}$, $\Sir_\SSC{ind}$, $G^A$. Output:
$\Sil(t,t')$ (and $G^<({\rm k};t,t')$, if required -- not shown). Main
differences with respect to $\Sir_\SSC{ind}(t,t')$ (see  Fig.5):\ \DIA\
non-iterative process (see text);\ \DIA\ the initial time $t'$ \ldots outermost
loop, the current time $t$ \ldots innermost loop. This "normal" order of loops
was reverted for $\Sir$. } \label{fig:flwch2}
\end{figure}

 \begin{figure}
 \caption{Imaginary part of the induced self-energy \Eq{eq:siri_expl_sum}
 as a function of both time variables. Insert: the same plot rotated to the
 view in the direction of the arrow (along the $t+t'$ axis). Vertical scale unit: eV/ps.}
 \label{fig:sirtt}
 \end{figure}

 \begin{figure}
 \caption{Band diagonal elements  \protect$|\ci\hbar\Gr_{bb}|^2$ for resonant wave vector
 ${\rm k}^{\times}$ as a function of $t$ and $t'$. Upper panel: $b=c$. Lower
 panel: $b=v$. In each panel, two surfaces are marked by dots, the full SCBA
 $\Gr$, and its approximation with $\Sir_\SSC{i}$ neglected. As an aid for eye,
 the exact surface has several lines $t' =\,$const marked by heavier points.
 The dot patterns permit to view the surfaces either in the $t,t'$ coordinates,
 or in the Wigner coordinates $t\pm t'$.}
 \label{fig:gr_two}
 \end{figure}

 \begin{figure}
 \caption{Test of the semigroup multiplicative property \Eq{eq:sg_prop}.
 Heavy line: $\ci\hbar {G}^R(t,-0.2)$.
 Thin lines: $\ci\hbar {G}^R(t,t'')\ci\hbar {G}^R(t'',-0.2)$ for $t>t''$.
 Quantity plotted: $|i\hbar
 G_{cc}|^2+|i\hbar G_{vc}|^2$ (upper panels) and $|i\hbar
 G_{vv}|^2+|i\hbar G_{cv}|^2$ (lower panels).
 The pulse envelope \Eq{eq:ext_field}  and the factorization times $t"$ (squares) shown at the
 top. The first, third, and fifth $t"$ were singled out also in Fig.
 \ref{fig:gr_two}.
 Left hand column: full SCBA, $\Sigma_{ind}(t,t')$ included.
 Right hand column: WW approximation, $\Sigma_{ind}(t,t')$ neglected. Here, the
 factorization is valid and the thin lines merge with the basic heavy plot.}
 \label{fig:sg}
 \end{figure}

\begin{figure}
\caption{Self-energy $\Sil(t,t')$ as a self-consistent result of eqs.
\Eq{eq:less_eq_def} and \Eq{eq:sili_expl_sum}. Upper panel: real part, lower
panel: imaginary part. Inserts: the same plot viewed in the direction of the
arrow (along the $t+t'$ axis). Units of $\Si$: eV/ps.}
\label{fig:sigma_less}
\end{figure}

\begin{figure}
\caption{The particle correlation function $\Gl_{cc}({\rm k};t,t')$
 for resonant detuning ${\rm k}={\rm k}^{\times}$.
 Upper panel: real part, lower panel:
 imaginary part. Inserts: view along the arrow --
 as in Fig.\ref{fig:sigma_less}. Units: $\hbar G$ is dimensionless}
\label{fig:g_less}
\end{figure}

\begin{figure}
\caption{$\Im G^<_{cc}({\rm k}^{\times};t,t')$ of the previous figure split
into its coherent (upper panel) and incoherent (lower panel) parts according to
\Eq{eq:g_less_dyson}.} \label{fig:g_less_coh}
\end{figure}

\begin{figure}
\caption{The coherent part $\Im \Gl_{cc,\SSC{coh}}({\rm k}^{\times};t,t')$ of
Fig. \protect\ref{fig:g_less_coh}
 (upper surface, coarse grid) contrasted with
the calculation, where $\Sir_\SSC{ind}(t,t')$ was neglected.(lower surface,
fine grid).} \label{fig:g_less_noind}
\end{figure}

\begin{figure}
\caption{Time diagonal  $\Im G^<_{cc}({\rm k}^{\times};t,t'=t)$. Thick line:
data of Fig.\ref{fig:g_less}.
 Thin line: the same quantity, but
$\Sir_\SSC{ind}$ neglected in computation.}
\label{fig:g_less_diag}
\end{figure}

\begin{figure}
\caption{One-electron density matrix decomposed into its band and $k$-vector
dependent components. Detuning $\Di$ equals to the excess energy $\ep_c({\rm
k})-\ep_c({\rm k}^{\times})$ measured with respect to the one photon resonance.
Normalization to one primitive cell, $\rho$ is dimensionless. The $\Di = 0$
profile identical with the thick line of Fig. \ref{fig:g_less_diag}}
\label{fig:rho_k}
\end{figure}

\begin{figure}
\caption{Total photoexcitation $\rho_{ab}^\SSC{tot}$, \EEq{eq:rho_tot}.
Diagonal panels: total band population per cell (thick line ... direct GF
computation; thin line ... coherent part only; dashed line ... computation
neglecting $\Sir_\SSC{ind}$). Conservation law
$\rho_{cc}^\SSC{tot}+\rho_{vv}^\SSC{tot}=1$: very good for direct GF
computation, poor in both other cases. Off-diagonal panels: The $vc$ corner:
the off-diagonal component $\rho_{vc}^\SSC{tot}$. The $cv$ corner: the same
quantity in semi-logarithmic form $\rho_{vc}^\SSC{tot}=R{\rm e}^{\ci \varphi}$.
Thick line \ldots $R$. Thin line \ldots $\varphi$. Dotted line \ldots pulse
envelope $\Phi$ scaled to coincide with $R$ at early times. Note the good fit.
All quantities shown are dimensionless.}
\label{fig:distr_tot}
\end{figure}

\begin{figure}
\caption{Power absorbed per primitive cell and averaged over the $\Omi$ cycle
... thick line; integral energy transfer between the pulse and one primitive
cell ... thin line Units: eV/ps for $\Upsilon_0 w$, eV for $\Upsilon_0 W$.}
\label{fig:joule}
\end{figure}
 \end{document}